\RequirePackage{rotating}
\documentclass[useAMS,usenatbib]{mnras}

\usepackage{ulem}
\usepackage{color}
\usepackage{graphics,graphicx}
\usepackage{times} 
\usepackage{amssymb}
\usepackage{amsmath}
\usepackage{natbib}
\usepackage{lscape}
\usepackage{url}
\usepackage{multirow}
\usepackage{ctable}
\usepackage{threeparttable}
\usepackage[labelfont=bf,labelsep=period]{caption}
\newif\ifAMStwofonts
\AMStwofontstrue

\def\asca{{\it ASCA}}
\def\ginga{{\it Ginga}}

\def\xmm{{\it XMM-Newton}}

\def\chandra{{\it Chandra}}
\def\swift{{\it Swift}}
\def\swiftng{{\it Neil Gehrels Swift Observatory}}
\def\integral{{\it INTEGRAL}}

\def\epicpn{{EPIC-pn}}
\def\epicmos1{{EPIC-MOS1}}
\def\epicmos2{{EPIC-MOS2}}
\def\epicmos{{EPIC-MOS}}

\def\nustar{{\it NuSTAR}}

\def\deg{$^{\circ}$}

\def\kmps{\hbox{$\rm\thinspace km~s^{-1}$}}
\def\pcmsq{\hbox{$\rm\thinspace cm^{-2}$}}
\def\H0{{km~s$^{-1}$~Mpc$^{-1}$}}

\def\kev{\hbox{\rm keV}}

\def\ergpcmsqps{\hbox{$\rm\thinspace erg~cm^{-2}~s^{-1}$}}

\def\ergps{\hbox{erg~s$^{-1}$}}
\def\ergcmps{\hbox{\rm erg~cm~s$^{-1}$}}

\def\pcmcub{\hbox{cm$^{-3}$}}
\def\msun{\hbox{$M_{\odot}$}}

\def\kte{$kT_{\rm{e}}$}

\def\lambdaCDM{$H_0$ = 70\,\H0, $\Omega_{\rm{M}} = 0.3$, $\Omega_{\Lambda} = 0.7$}

\def\addascaspec{\textsc{addascaspec}}

\def\flx2xsp{\textsc{flx2xsp}}

\def\nustardas{\textsc{nustardas}}
\def\nupipeline{\textsc{nupipeline}}
\def\nuproducts{\textsc{nuproducts}}

\def\sas{\textsc{sas}}
\def\xmmselect{\textsc{xmmselect}}

\def\epchain{\textsc{epchain}}
\def\emchain{\textsc{emchain}}

\def\rmfgen{\textsc{rmfgen}}
\def\arfgen{\textsc{arfgen}}

\def\chisq{{$\chi^{2}$}}

\def\xspec{\hbox{\textsc{xspec}}}
\def\xspecv{\hbox{\textsc{xspec} v12.10.1s}}

\def\xstar{\textsc{xstar}}

\def\tbabs{\textsc{tbabs}}

\def\xillver{\textsc{xillver}}
\def\xillvercp{\textsc{xillver\_cp}}

\def\relline{\textsc{relline}}
\def\relxill{\textsc{relxill}}

\def\relxilllpioncp{\textsc{relxilllp\_ion\_cp}}
\def\relxilllpdcp{\textsc{relxilllp\_d\_cp}}
\def\pexrav{\textsc{pexrav}}

\def\borus{\textsc{borus}}

\def\nthcomp{\textsc{nthcomp}}
\def\cutoffpl{\textsc{cutoffpl}}
\def\mekal{\textsc{mekal}}

\def\fexxv{\hbox{\rm Fe\,{\small XXV}}}
\def\fexxvi{\hbox{\rm Fe\,{\small XXVI}}}

\def\eg{{\it e.g.}}

\def\ie{{\it i.e.~\/}}

\def\la{\mathrel{\hbox{\rlap{\hbox{\lower4pt\hbox{$\sim$}}}{\raise2pt\hbox{$<$}}}}}
\def\ga{\mathrel{\hbox{\rlap{\hbox{\lower4pt\hbox{$\sim$}}}{\raise2pt\hbox{$>$}}}}}

\def\d25{D$_{25}$}
\def\nh{{$N_{\rm H}$}}

\def\.25{0.25 keV\thinspace}

\def\lbol{\rm $L_{\rm bol}$}
\def\kbol210{\rm $\kappa_{2-10}$}

\def\rg{$R_{\rm{G}}$}
\def\rh{$R_{\rm{H}}$}

\def\Rfrac{$R_{\rm{frac}}$}

\def\eso{ESO\,033--G002}
\def\nwalker{60}
\def\nstep{30,000}
\def\nburn{5,000}
\def\totchain{1,500,000}
\newcommand{\Cahk}{Ca H+K}
\newcommand{\Mgb}{Mg\,{\textsc{i}}}
\newcommand{\caii}{Ca\,{\textsc{ii}}}

\title[Extreme Reflection in \eso]{Extreme Relativistic Reflection in the Active Galaxy \eso}

\author[D. J. Walton, et al.]
{\parbox{7.in}{D. J. Walton$^{1}$\thanks{E-mail: dwalton@ast.cam.ac.uk},
M. Balokovi\'c$^{2,3}$,
A. C. Fabian$^{1}$,
L. C. Gallo$^{4}$,
M. Koss$^{5}$,
E. Nardini$^{6,7}$, \\
C. S. Reynolds$^{1}$,
C. Ricci$^{8,9,10}$,
D. Stern$^{11}$,
W. N. Alston$^{12}$,
T. Dauser$^{13}$,
J. A. Garc\'ia$^{14,13}$, \\
P. Kosec$^{15,1}$,
M. T. Reynolds$^{16}$,
F. A. Harrison$^{14}$,
J. M. Miller$^{16}$
\\[0.25cm]
\footnotesize
$^{1}$ \it{Institute of Astronomy, University of Cambridge, Madingley Road, Cambridge CB3 0HA, UK} \\
$^{2}$ \it{Yale Center for Astronomy \& Astrophysics, 52 Hillhouse Avenue, New Haven, CT 06511, USA} \\
$^{3}$ \it{Department of Physics, Yale University, P.O. Box 2018120, New Haven, CT 06520, USA} \\
$^{4}$ \it{Department of Astronomy and Physics, Saint Mary's University, 923 Robie Street, Halifax, NS B3H 3C3, Canada} \\
$^{5}$ \it{Eureka Scientific, 2452 Delmer Street Suite 100, Oakland, CA 94602-3017, USA} \\
$^{6}$ \it{Dipartimento di Fisica e Astronomia, Universit\`a di Firenze, via G. Sansone 1, I-50019 Sesto Fiorentino, Firenze, Italy} \\
$^{7}$ \it{INAF -- Osservatorio Astrofisico di Arcetri, Largo Enrico Fermi 5, I-50125 Firenze, Italy} \\
$^{8}$ \it{N\'ucleo de Astronom\'ia de la Facultad de Ingenier\'ia, Universidad Diego Portales, Av. Ej\'ercito Libertador 441, Santiago, Chile} \\
$^{9}$ \it{Kavli Institute for Astronomy and Astrophysics, Peking University, Beijing 100871, China} \\
$^{10}$ \it{George Mason University, Department of Physics \& Astronomy, MS 3F3, 4400 University Drive, Fairfax, VA 22030, USA} \\
$^{11}$ \it{Jet Propulsion Laboratory, California Institute of Technology, Pasadena, CA 91109, USA} \\
$^{12}$ \it{European Space Agency (ESA), European Space Astronomy Centre (ESAC), E-28691 Villanueva de la Ca\~nada, Madrid, Spain} \\
$^{13}$ \it{Dr. Karl Remeis-Observatory and Erlangen Centre for Astroparticle Physics, Sternwartstr.~7, D-96049 Bamberg, Germany} \\
$^{14}$ \it{Cahill Center for Astronomy and Astrophysics, California Institute of Technology, Pasadena, CA 91125, USA} \\
$^{15}$ \it{MIT Kavli Institute for Astrophysics and Space Research,70 Vassar Street, Cambridge, MA 02139, USA} \\
$^{16}$ \it{Department of Astronomy, University of Michigan, 1085 S. University, Ann Arbor, MI 48109, USA} \\
}}
\date{}

\begin{document}
\pagerange{\pageref{firstpage}--\pageref{lastpage}}
\maketitle
\label{firstpage}

\begin{abstract}
We present the first high signal-to-noise broadband X-ray spectrum of the radio-quiet
type-2 Seyfert \eso, combining data from \xmm\ and \nustar. The nuclear X-ray
spectrum is complex, showing evidence for both neutral and ionised absorption, as
well as reflection from both the accretion disc and more distant material, but our 
broadband coverage allows us to disentangle all of these different components. The 
total neutral column during this epoch is $N_{\rm{H}} \sim (5-6) \times 10^{22}$\,\pcmsq, 
consistent with the optical classification of \eso\ as a type-2 Seyfert but not so large as
to prevent us from robustly determining the properties of the innermost accretion flow.
The ionised absorption -- dominated by lines from \fexxv\ and \fexxvi\ -- reveals a
moderately rapid outflow ($v_{\rm{out}} \sim 5400$\,\kmps) which has a column 
comparable to the neutral absorption. We find the disc reflection from the innermost
regions to be extreme, with a reflection fraction of \Rfrac\ $\sim$ 5. This requires
strong gravitational lightbending and, in turn, both an extremely compact corona
(within $\sim$2\,\rg\ of the black hole) and a rapidly rotating black hole ($a^* > 0.96$).
Despite this tight size constraint, with a temperature of $kT_{\rm{e}} = 40-70$\,keV
the X-ray corona in \eso\ appears similar to other AGN in terms of its placement in
the compactness-temperature plane, consistent with sitting close to the limit
determined by runaway pair production. Finally, combining X-ray spectroscopy,
timing and updated optical spectroscopy, we also estimate the mass of the black
hole to be $\log[M_{\rm{BH}} / \msun] \sim 7.0 - 7.5$.
\end{abstract}


\section{Introduction}

The primary source of X-ray emission in active galactic nuclei (AGN) is the corona,
a hot plasma of electrons that is widely expected to Compton up-scatter thermal
emission from the accretion disc -- which typically peaks in the EUV band for AGN
(\eg\ \citealt{Shakura73, Elvis94}) -- up to higher energies. This produces a
powerlaw-like X-ray continuum with a high-energy cutoff set by the temperature of the
coronal electrons (\eg\ \citealt{Haardt91}). Although the fine details of the corona are
still being explored, observations show that its emission can be highly and rapidly
variable (\eg\ \citealt{Fabian13iras}), indicating it must be a fairly compact structure.
Independent constraints on the size of the X-ray source from microlensing (\eg\
\citealt{Chartas09, Dai10, MacLeod15}), X-ray occultation events (\eg\
\citealt{Risaliti09a, Gallo21}) and X-ray reverberation (\eg\ \citealt{Kara16,
Alston20iras}) confirm this, implying a typical size scale of $\sim$10\,\rg\ (where \rg\
= $GM_{\rm{BH}}/c^2$ is the gravitational radius, and $M_{\rm{BH}}$ is the black
hole mass).

In addition to this Comptonised continuum, the X-ray spectra of AGN also typically
reveal evidence for significant reprocessing of this emission. Most notably, the X-ray
emission from the corona will irradiate the surface of the accretion disc, resulting in a
series of emission lines (via fluorescence) and a characteristic backscattered
continuum at high energies (peaking at $\sim$20--30\,keV; \eg\ \citealt{George91}).
Because of its high cosmic abundance and high fluorescent yield, the strongest line is 
typically iron K$\alpha$ at 6.4--6.97\,keV (depending on the ionisation state). Owing
to the rapid orbital motions and the strong gravity close to the black hole, special and
general relativistic effects will broaden and skew line emission from the inner disc into
a characteristic `diskline' profile which has a strongly extended red wing (\eg\
\citealt{Fabian89, kdblur}). This `reflected' emission is of critical importance as it
carries information on the innermost accretion geometry, including both the extent of
the innermost accretion disc -- which is in turn set by the spin of the black hole
(\citealt{Bardeen72}) -- and the geometry of the corona (\eg\ \citealt{Martocchia00,
lightbending, Wilkins12}). Spin measurements are of particular interest for AGN, as
they provide a window into the (recent) growth history of the supermassive black
holes that power them, revealing whether this likely occurred via accretion or
mergers (\eg\ \citealt{King06, Sesana14, Fiacconi18}). Black hole spin may also be a
key ingredient for powering the relativistic jets these systems can launch
(\eg\ \citealt{BZ77}).

Broadband spectroscopy is key to understanding all of these issues. For AGN, the
best quantified method for constraining spin in individual AGN is via the study of
the reflection from the inner disc, and the most robust constraints come when both
the iron line and the reflected continuum can be studied simultaneously (\eg\
\citealt{Risaliti13nat, Walton14, Walton20iras, Marinucci14}). This is also naturally
the case for using this emission to place geometric constraints on the corona.
Furthermore, sensitive hard X-ray coverage (\ie\ $E > 10$\,keV) is necessary to
provide robust constraints on the coronal temperature (\eg\ \citealt{Brenneman14a,
Malizia14, Matt15, Balokovic15, Buisson18cutoff}) and understand its plasma
properties.

\eso\ (aka IRAS\,04575--7537) is a nearby ($z$ = 0.0181) radio-quiet Seyfert-2 active
galaxy (\citealt{deGrijp92, Vignali98}) that is X-ray bright, detected in the hard X-ray 
surveys (\citealt{BAT9m, 3ISGRI}) undertaken by both the \swiftng\ (hereafter \swift;
\citealt{SWIFT}) and \integral\ (\citealt{INTEGRAL}). Despite this, however, it has
received little dedicated observational attention to date.  A fairly deep observation with
\asca\ (\citealt{ASCA}) and snapshots with \ginga\ (\citealt{GINGA}) and \chandra\
(\citealt{CHANDRA}) revealed evidence for moderate X-ray obscuration ($N_{\rm{H}}
\sim 10^{22}$\,\pcmsq; \citealt{Smith96, Vignali98, Marchesi17}), broadly consistent
with its optical classification. In addition, a short observation performed with \nustar\
(\citealt{NUSTAR}) as part of its extragalactic survey campaign showed evidence for
very strong contribution from reflected emission (reflection strength of $R \sim 2-3$;
\citealt{Balokovic17phd, Lanz19, Panagiotou19}). This is among the largest reflection
contributions seen among local, Compton-thin AGN (\citealt{Walton13spin}), and may
imply an extreme accretion geometry in which strong lightbending suppresses the
direct emission from the primary X-ray source, leaving a spectrum dominated by
reflection from the innermost accretion disc (\eg\ \citealt{lightbending, Fabian12,
Parker14mrk, Dauser14}).

Here we present new broadband X-ray observations of \eso\ taken with \xmm\
(\citealt{XMM}) and \nustar, from which we are able to place tight constraints on the
geometry of the X-ray source via the reflected emission, providing the first constraints
on the spin of this SMBH, as well as updated constraints on the coronal temperature
in this source.

\begin{figure*}
\begin{center}
\hspace*{-0.4cm}
\rotatebox{0}{
{\includegraphics[width=235pt]{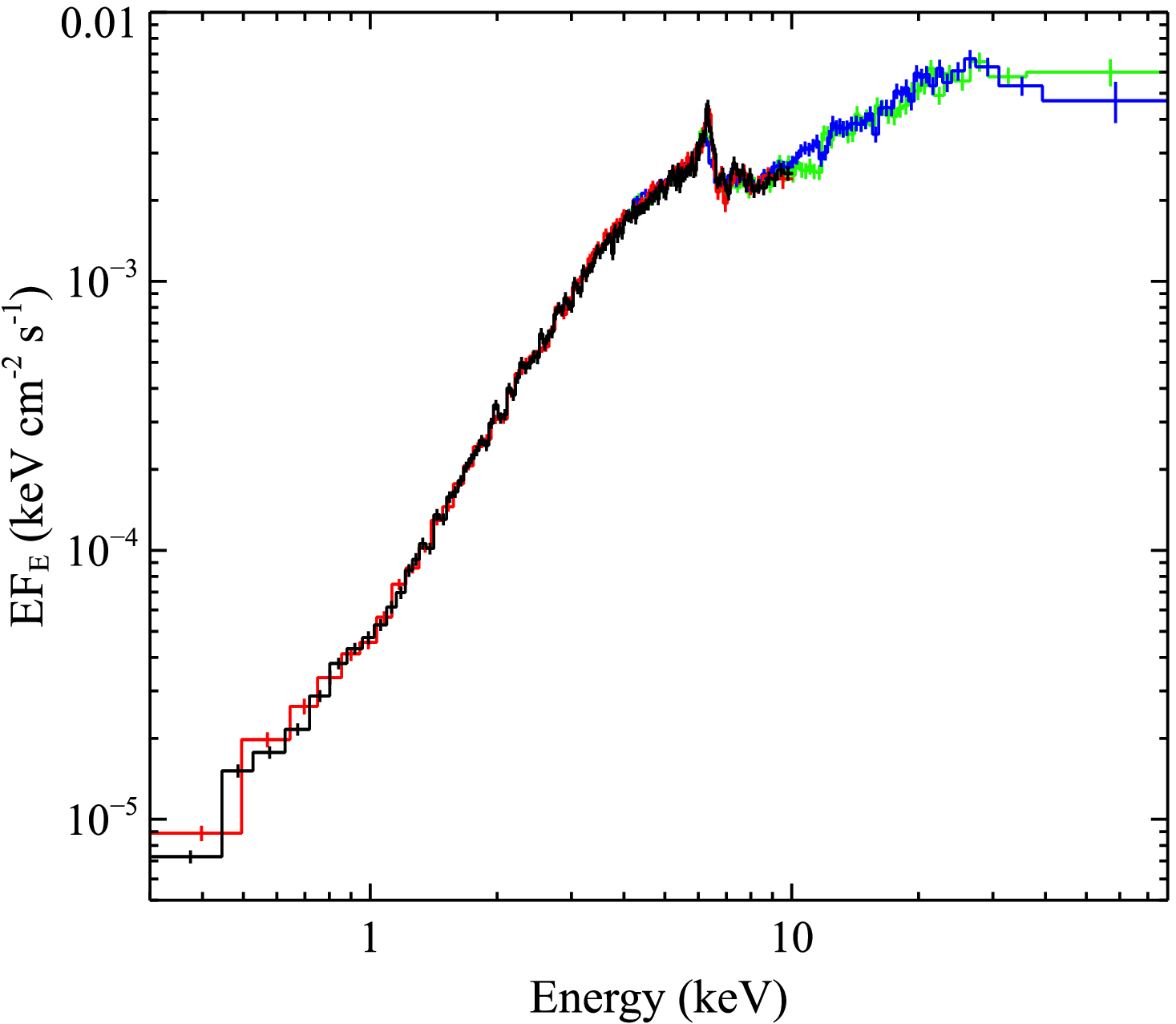}}
}
\hspace*{0.6cm}
\rotatebox{0}{
{\includegraphics[width=235pt]{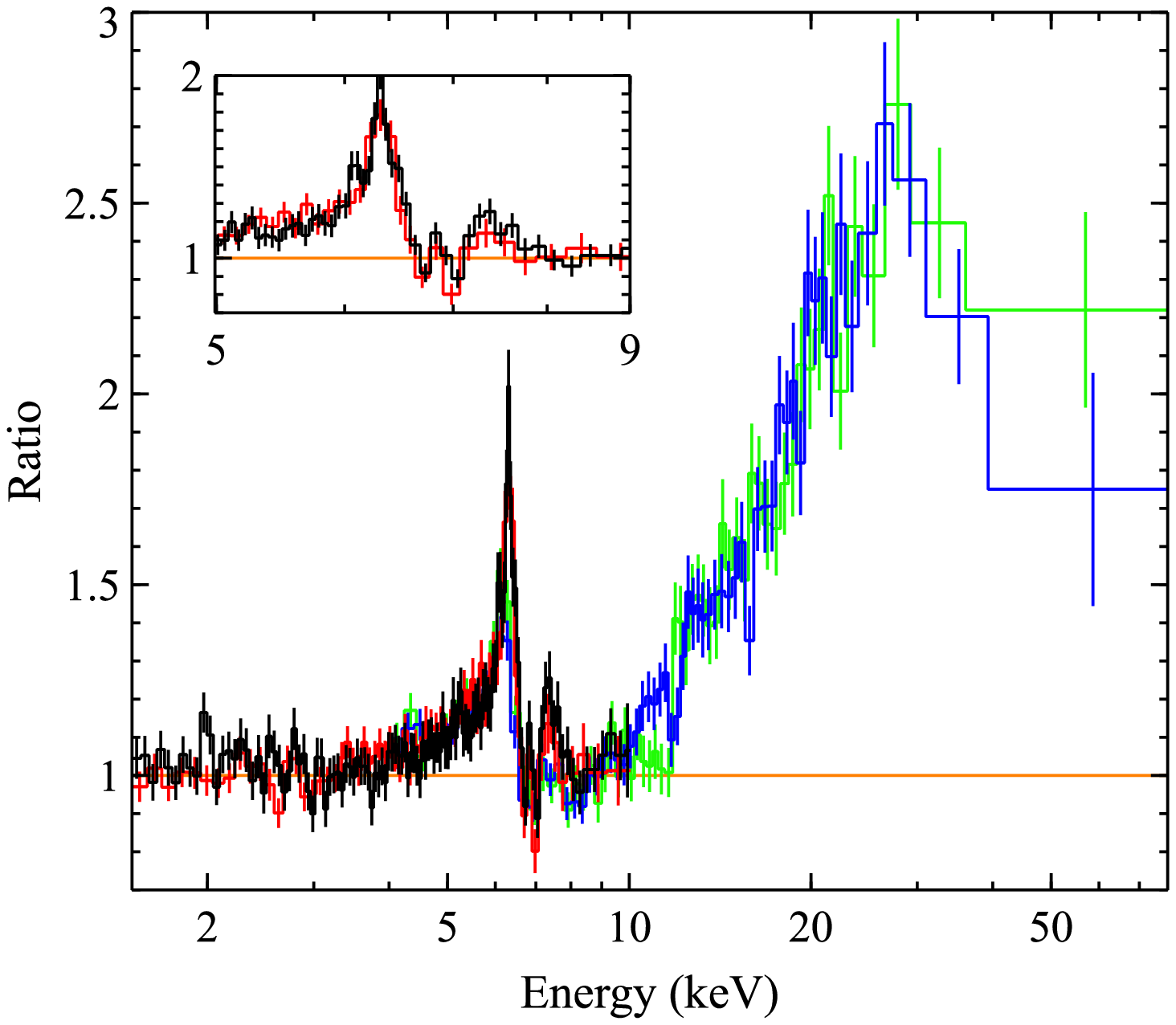}}
}
\end{center}
\vspace*{-0.3cm}
\caption{
Left panel: the broadband \xmm+\nustar\ spectrum of \eso\ (after being unfolded
through a model that is constant with energy). The \xmm\ data are shown in black and
red (\epicpn\ and \epicmos, respectively) and the \nustar\ data are shown in blue and
green (FPMA and FPMB, respectively). These new observations show the source
remains moderately absorbed, as also seen in the archival \ginga, \asca\ and
\chandra\ data (\citealt{Smith96, Vignali98, Marchesi17}). Right panel: residuals to a 
simple \cutoffpl\ continuum, modified by a partially covering neutral absorber, and 
applied to the broadband data over the 1.5--4, 7--10 and 50--78\,\kev\ energy ranges. 
As with  other similar cases with coordinated \xmm+\nustar\ observations (\eg\ 
\citealt{Risaliti13nat, Marinucci14, Parker14mrk, Walton14, Walton18, Walton19ufo,
Walton20iras}), the key signatures of relativistic disc reflection are seen: 
relativistically broadened iron line at $\sim$6\,\kev\ and a strong Compton hump
peaking at $\sim$30\,keV. In addition, a clear narrow core to the iron emission is also
seen at 6.4\,keV, indicating reprocessing by more distant material, and a pair of
absorption lines from ionised iron (\fexxv\ and \fexxvi) are also clearly present at
$\sim$7\,keV (as shown in the inset). Note that the data in all panels have been
rebinned for visual purposes.}
\label{fig_spec}
\end{figure*}

\section{Observations and Data Reduction}
\label{sec_red}

We undertook a long coordinated observation of \eso\ with both \xmm\ and \nustar\ in
June 2020. The following sections describe our reduction of these data.

\begin{table}
  \caption{Details of the coordinated \xmm\ and \nustar\ observations of \eso\
  considered in this work.}
%\vspace{-0.5cm}
\begin{center}
\begin{tabular}{c c c c c}
\hline
\hline
\\[-0.25cm]
Mission & OBSID & Start & Good \\
& & Date & Exposure\tmark[a] \\
\\[-0.3cm]
\hline
\hline
\\[-0.2cm]
\xmm\ & 0863050201 & 2020-06-09 & 109/125 \\
\\[-0.3cm]
\nustar\ & 60601002002 & 2020-06-09 & 172 \\
\\[-0.2cm]
\hline
\hline
\\[-0.15cm]
\end{tabular}
\label{tab_obs}
\end{center}
\vspace{-0.3cm}
$^{a}$ Exposures are given in ks, and for \xmm\ are listed for the \epicpn/MOS detectors
after periods of high background have been removed.
%\vspace*{0.3cm}
\end{table}

\subsection{\textit{NuSTAR}}

The \nustar\ data for both focal plane modules (FPMA and FPMB) were reduced with
the \nustar\ Data Analysis Software (\nustardas) v1.9.2, and \nustar\ calibration
database v20190627. The unfiltered event files were cleaned with \nupipeline,
using the standard depth correction to reduce the internal high-energy background.
Passages through the South Atlantic Anomaly were excluded from consideration
using the following settings: {\textsc{saa = optimized}} and {\textsc{tentacle = yes}}.
Source products and their associated instrumental response files were then extracted
from these cleaned event lists using \nuproducts\ and a circular region of radius
90$''$, while the background was estimated from a larger region of blank sky on the
same chip as \eso. Note that in this case, all of the on-source data is in the standard
`science' mode (mode 1), and there is no indication that this observation was affected
by the recently identified tear in the thermal blanket (\citealt{NuSTARmli}). We use
the \nustar\ data over the 4--78\,keV band, and the spectra are rebinned to a
minimum signal-to-noise (S/N) of 5 per energy bin.

\subsection{\textit{XMM-Newton}}

The \xmm\ data were reduced using the \xmm\ Science Analysis System (\sas\
v18.0.0). Following standard procedures, the raw observation files were cleaned
using \epchain\ and \emchain\ for the \epicpn\ and \epicmos\ detectors, respectively
(\citealt{XMM_PN, XMM_MOS}). Source products were extracted from the cleaned 
event files with \xmmselect\ from a circular region of radius 35$''$, using only single
and double patterned events for \epicpn\ ({\small PATTERN}\,$\leq$\,4) and single to
quadruple patterned events for \epicmos\ ({\small PATTERN}\,$\leq$\,12). Periods of
high background were filtered out, and similar to the \nustar\ data, the remaining
background contribution was again estimated from a larger region of blank sky on
the same chip as \eso. Instrumental response files were generated using \arfgen\
and \rmfgen\ for each of the EPIC detectors. After performing the reduction
separately for the \epicmos1\ and \epicmos2\ detectors, we produced a combined 
\epicmos\ spectrum using \addascaspec. We use the \xmm\ data over the
0.3--10.0\,keV band, and the spectra are again rebinned to a minimum S/N of 5 per
energy bin.

\section{Broadband Spectroscopy}
\label{sec_spec}

\begin{figure}
\begin{center}
\hspace*{-0.3cm}
\rotatebox{0}{
{\includegraphics[width=245pt]{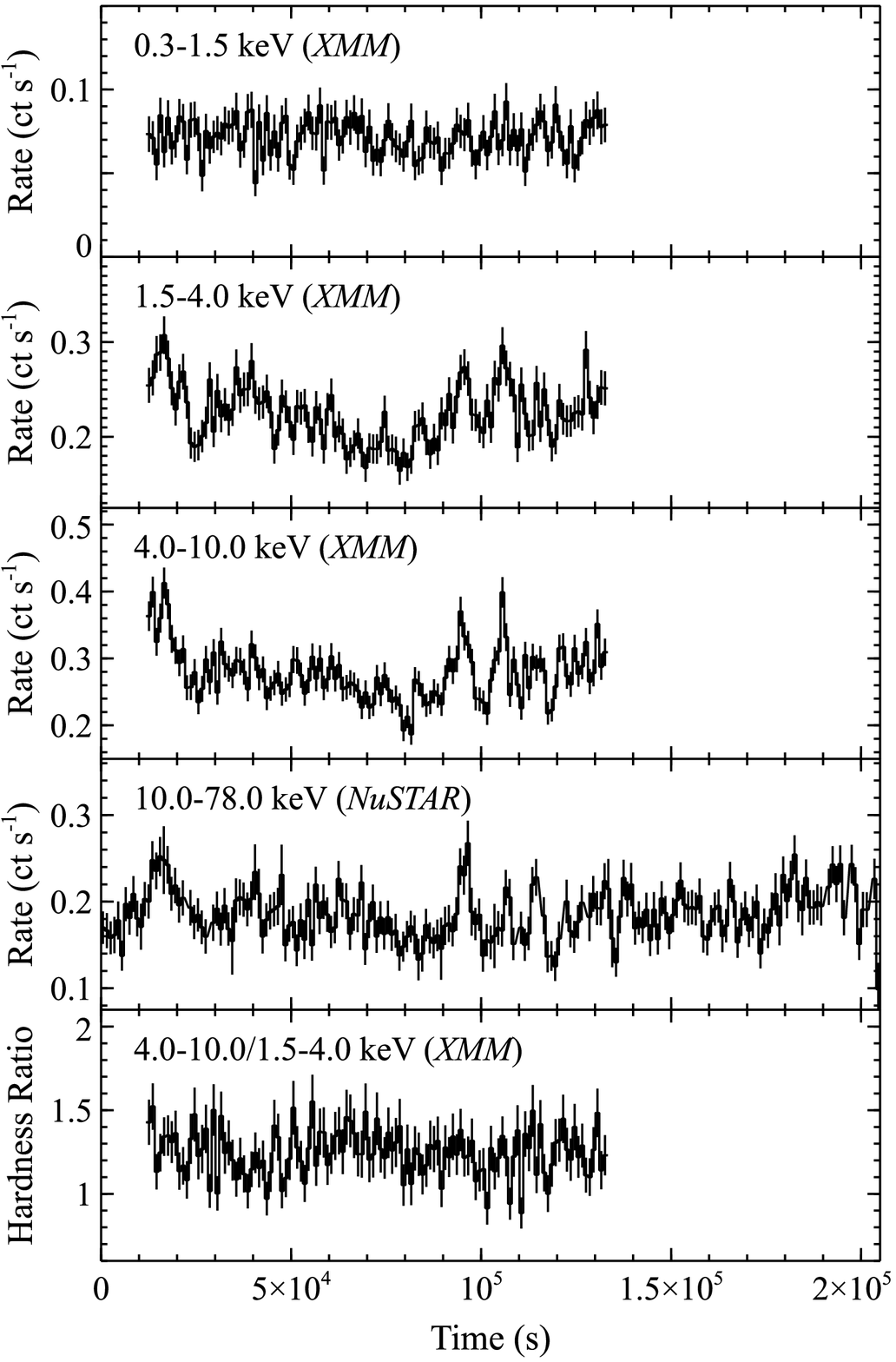}}
}
\end{center}
\vspace*{-0.3cm}
\caption{Energy-resolved lightcurves of \eso, shown with 1\,ks bins (\epicpn\ data for
\xmm, combined FPMA+FPMB data for \nustar). The top three panels show the
0.3--1.5\,keV, 1.5--4.0\,keV and 4.0--10.0\,keV bands from \xmm, and the fourth
panel shows the 10--78\,keV band from \nustar, respectively. Clear flux variability is
seen in the latter three bands, where the direct emission from the nucleus dominates.
The bottom panel shows the hardness ratio between the latter two highest energy
bands from \xmm; there is little evidence for strong spectral variability associated with
the observed flux variability.
}
\label{fig_lc}
\end{figure}

The time-averaged broadband spectrum of \eso\ is shown in Figure \ref{fig_spec}
(left panel). In a qualitative sense, these data confirm the moderately obscured
nature of the source revealed by the archival \ginga, \asca\ and \chandra\ data
(\citealt{Smith96, Vignali98, Marchesi17}), with the direct AGN emission dominating
above $\sim$1.5\,keV. In obscured AGN, the softest X-rays are typically dominated
by scattered nuclear flux and diffuse plasma emission (\eg\ \citealt{Bianchi05mrk3,
Miniutti07, Winter08}). In order to confirm that this is also the case in \eso\ as well,
we extracted lightcurves in a variety of different energy bands. Flux variability
associated with the central AGN is clearly seen above $\sim$1.5\,keV, while the
softer X-ray data do not show similar variations (see Figure \ref{fig_lc}), instead
remaining roughly constant (as expected for scattered/plasma emission from large
scales). Furthermore, although a detailed spatial analysis is beyond the scope of
this work, we also note that the archival \chandra\ imaging does show potential
hints of low-level extended X-ray  emission around the nucleus.

As has become standard, in order to highlight the other key spectral features we also
show the data/model ratio of the combined \xmm+\nustar\ dataset to an absorbed
powerlaw continuum in Figure \ref{fig_spec} (right panel), fit to the 1.5--4, 7--10 and
50--78\,keV bands where the primary AGN continuum would be expected to
dominate; model fits are performed with \xspecv\ (\citealt{xspec}) using \chisq\
minimisation. We use the \tbabs\ neutral absorption code (\citealt{tbabs})\footnote{As
recommended, we use the cross-sections of \cite{Verner96} for the absorption.
However, following our previous work (\citealt{Walton18, Walton19ufo, Walton20iras}),
we combine these with the solar abundance set of \cite{Grevesse98} for internal
self-consistency with both the \xillver\ reflection model (\citealt{xillver}) and the \xstar\
photoionisation code (\citealt{xstar}), which are heavily utilized throughout our
analysis; see Section \ref{sec_mod}.}, and we allow for both fully and partially
covering absorption. In addition to the absorbers associated with \eso, all models
presented in this work also include neutral absorption from the Galactic column (also 
using \tbabs; $N_{\rm{H,Gal}} = 8.95 \times 10^{20}$\,\pcmsq, \citealt{NH2016}),
As is standard, we also allow cross-normalisation constants to float between the
different detectors throughout this work (taking \epicpn\ as our reference point), in
order to account for both calibration differences and the fact that the \nustar\
observation spans a longer duration than \xmm\ does; these are always within
$\sim$15\% of unity, broadly as expected given the current level of cross-calibration
(\eg\ \citealt{NUSTARcal}) and the level of variability observed (see Figure
\ref{fig_lc}). The column density of the fully covering absorber is $N_{\rm{H}} \sim 2
\times 10^{22}$\,\pcmsq, similar to the previous observations of \eso\ highlighted
above, the column density and covering factor for the partially covering absorber are
$N_{\rm{H}} \sim 10^{23}$\,\pcmsq\ and $C_{\rm{f}} \sim 0.8$, respectively, and the
photon index is $\Gamma \sim 2$, fairly typical for AGN (\eg\ \citealt{Ricci17}). A
strong, broad feature is clearly present in the iron K bandpass, along with a strong
hard excess that peaks at $\sim$30\,keV. Both features are key characteristics of
strong reflection from the inner accretion disc.

A narrow core to the iron emission is also seen at $\sim$6.4\,keV, indicating further
reprocessing by more distant material, and a pair of absorption lines associated with
highly ionised iron are also seen at $\sim$7\,keV. \eso\ therefore shows similar
spectral complexity to a number of other famous and well-studied AGN, \eg\
NGC\,1365 (\citealt{Risaliti05b, Risaliti09a, Risaliti13nat, Walton14, Rivers15}),
IRAS\,13197--1627 (\citealt{Dadina04, Miniutti07, Walton18}), among others.
Modelling the 1.5--10.0\,keV band with the above continuum, three Gaussians for
the narrow features, and a \relline\ component (\citealt{relconv}) for the broad iron
emission, we find the latter has an extremely large equivalent width of $EW =
500^{+700}_{-90}$\,eV, the narrow core has a much more modest equivalent width
of $EW = 65 \pm 20$\,eV (note that we quote parameter uncertainties at the 90\%
confidence level for one interesting parameter). For the \relline\ component, we limit
the line energy to 6.4--6.97\,keV in the rest-frame of \eso, corresponding to neutral
and hydrogen-like iron, and find that the line energy can have any value in this
range. Assuming the disc extends into the innermost stable circular orbit, the
best-fit spin and inclination are $a^* \gtrsim 0.8$ and $i \sim 50$\deg. The \relline\
model allows for a broken powerlaw emissivity profile, but we assume a single
powerlaw for simplicity at this stage, \ie $\epsilon(r) \propto r^{-q}$, and find a fairly
steep index of $q \sim 5$. However we stress that these values should only be
considered very preliminary, as a proper characterisation of the relativistic blurring
requires consideration of the full reflection spectrum (performed below),
self-consistently treating both the iron emission and the corresponding iron edge
at higher energies.

The ionised absorption lines both have $EW \sim 100$\,eV (formally,
$EW_{\rm{XXV}} = 100 \pm 20$\,eV and $EW_{\rm{XXVI}} = 105 \pm 30$\,eV)
and line widths of $\sigma = 65 \pm 25$\,eV (assuming the two absorption lines
have the same width). This width roughly corresponds to a velocity broadening of
$\sim$3000\,\kmps, fairly typical for such absorption in other systems (\eg\
\citealt{Risaliti05b, Walton18}). The line energies in the rest-frame of \eso\ are $6.79
\pm 0.02$ and $7.09 \pm 0.03$\,keV, respectively, corresponding to a line-of-sight
outflow velocity of $\sim$5500\,\kmps.

\subsection{General Model Setup}
\label{sec_mod}

Although flux variability is clearly seen, there is limited evidence for significant
spectral variability even above 2\,keV (where the direct nuclear emission dominates;
see Figure \ref{fig_lc}), and so we focus our detailed spectral analysis on modelling
the time-averaged broadband \xmm+\nustar\ spectrum shown above (Figure
\ref{fig_spec}). We construct a spectral model for \eso\ that includes the intrinsic
emission from the central AGN, which consists of the primary X-ray continuum and
the associated relativistic reflection from the inner accretion disc, highly ionised
absorption, a complex neutral absorber, and reprocessing by more distant material.
Our approach to constructing this model is broadly similar to the recent broadband
analyses of IRAS\,13197-1627, IRAS\,00521-7054 and IRAS\,09149-6206 presented
by \cite{Walton18, Walton19ufo, Walton20iras}, respectively, all of which exhibit
similarly complex spectra to \eso.

Following these works, we model the disc reflection with the \relxill\ family of models,
(v1.3.10; \citealt{relxill, Dauser14}). We assume a lamppost geometry for the disc
reflection throughout our analysis. Although this is a somewhat simplistic geometry,
as discussed in previous works it allows for a physical interpretation for the reflection
fraction, \Rfrac\ (see \citealt{relxill_norm}), and also enables us to exclude
non-physical regions of parameter space (\eg\ a very steep radial emissivity profile
and a non-rotating black hole). As discussed below (Section \ref{sec_fit}), we
consider two of the available variants of the \relxill\ model that are potentially
relevant to \eso\ in our analysis. In these models, the ionising continuum is assumed
to be an  \nthcomp\ thermal Comptonisation model (\citealt{nthcomp1, nthcomp2}),
characterised by the photon index and electron temperature ($\Gamma,
kT_{\rm{e}}$; note that this emission is also included in the \relxill\ models). The
other key free parameters common to these models are the spin of the black hole
($a^*$), the height (above the spin axis) of the X-ray source ($h$) and the inclination, 
ionisation state and iron abundance of the accretion disc ($i$, $\xi_{\rm{in}}$ and 
$A_{\rm{Fe}}$, respectively; the rest of the elements included in the \xillver/\relxill\ 
models are assumed to have solar abundances). The ionisation parameter follows
the standard definition: $\xi = L_{\rm{ion}}/nR^{2}$, where $L_{\rm{ion}}$ is the
ionising luminosity (integrated over the 0.1--1000\,keV bandpass in \relxill), $n$ is
the density of the disc, and $R$ is the distance to the X-ray source; this is allowed to 
vary over the range $\log[\xi/(\rm{erg}~\rm{cm}~\rm{s}^{-1})]$ = 0--4, as appropriate
for AGN (\eg\ \citealt{Ballantyne11, Walton13spin}). We assume that the accretion
disc extends down to the innermost stable circular orbit (ISCO), and set the outer
radius of the disk to the maximum value allowed by the model (1000\,\rg).

We also consider two different models for the distant reflection: the \xillver\ model,
which assumes a simple slab geometry for the reprocessor, and the \borus\ model,
which assumes a torus-like geometry (formally the geometry is spherical with
conical polar cutouts). As part of the same broad family of models, \xillver\ naturally
shares most of its parameters with \relxill; for this component we assume the
material is essentially neutral ($\log[\xi/(\rm{erg}~\rm{cm}~\rm{s}^{-1})]$ = 0, the
lowest allowed value). For the  \borus\ model, we also use the variant that assumes
an \nthcomp\ continuum (\textsc{borus12}; \citealt{borus12}). The additional free
parameters in \borus\ are the column density through the torus, and the solid angle
it subtends; the reprocessing gas is assumed to be neutral in the \borus\ model.
Ultimately, though, we find consistent results regardless of which of these models
we use, and the additional free parameters in \borus\ are poorly constrained, so
we present the results with the \xillver\ model for simplicity 

For the neutral absorption associated with \eso, we again combine a partially
covering absorber with a fully covering component, continuing to use the \tbabs\
absorption code. The ionised absorption is instead modelled with \xstar, using a grid
of pre-calculated absorption models. These are computed assuming a generic
ionising continuum of $\Gamma = 2$ -- roughly similar to a typical AGN continuum
(\citealt{Ricci17}) -- to allow for broader applicability, and a velocity broadening of
3000\,\kmps\ (as found above); the key free parameters are the ionisation
parameter, column density, outflow velocity and iron abundance of the absorbing
gas; all other elements are assumed to have solar abundances. Note that the
bandpass for the ionising luminosity is defined to be 1--1000\,Ry (\ie 13.6\,eV --
13.6\,keV) in \xstar. The assumed $\Gamma = 2$ continuum is also similar to the
simple fit discussed previously, but even if the true photon index differs slightly from
this assumption, this will only result in mild differences in the ionisation parameter
inferred (see \eg\ \citealt{Walton20iras}). We also investigated allowing the neutral
absorber to be partially ionised, also using \xstar, but find that this offers no
improvement in the fit; the ionisation parameter tends towards neutral absorption
and the rest of the results are unchanged.

Finally, for the softest emission, we approximate this with a combination of a 
second \nthcomp\ continuum and a \mekal\ plasma component (\citealt{mekal1,
mekal2}) to represent scattered nuclear flux and diffuse plasma emission,
respectively, in order to allow for both possible contributions. The latter is an
optically-thin plasma that is simply  characterised by a plasma temperature, $kT$.

The model is set up such that the absorbers associated with \eso\ act only on the
emission from the central regions (\ie\ the \relxill\ component); the distant reflection
and the low-energy scattered/diffuse components are only subject to the Galactic
column. For self-consistency, we assume the same iron abundance for all the
different model components associated with the emission from the AGN in \eso,
and we also link all of the relevant continuum parameters among the different
model components ($\Gamma$, \kte), after accounting for the appropriate
gravitational redshift ($z_{\rm{grav}}$) implied by $a^*$ and $h$, since the lamppost
variants of \relxill\  evaluate the properties of the X-ray source in its rest frame. The
full \xspec\  expression for our basic model setup is as follows: \tbabs$_{\rm{Gal}}$
$\times$ $($\mekal\ + \nthcomp$_{\rm{scat}}$ + \xillver\ + $($\tbabs$_{\rm{full}}$
$\times$  \tbabs$_{\rm{part}}$ $\times$ \xstar\ $\times$ \relxill$))$.

\begin{figure}
\begin{center}
\hspace*{-0.35cm}
\rotatebox{0}{
{\includegraphics[width=235pt]{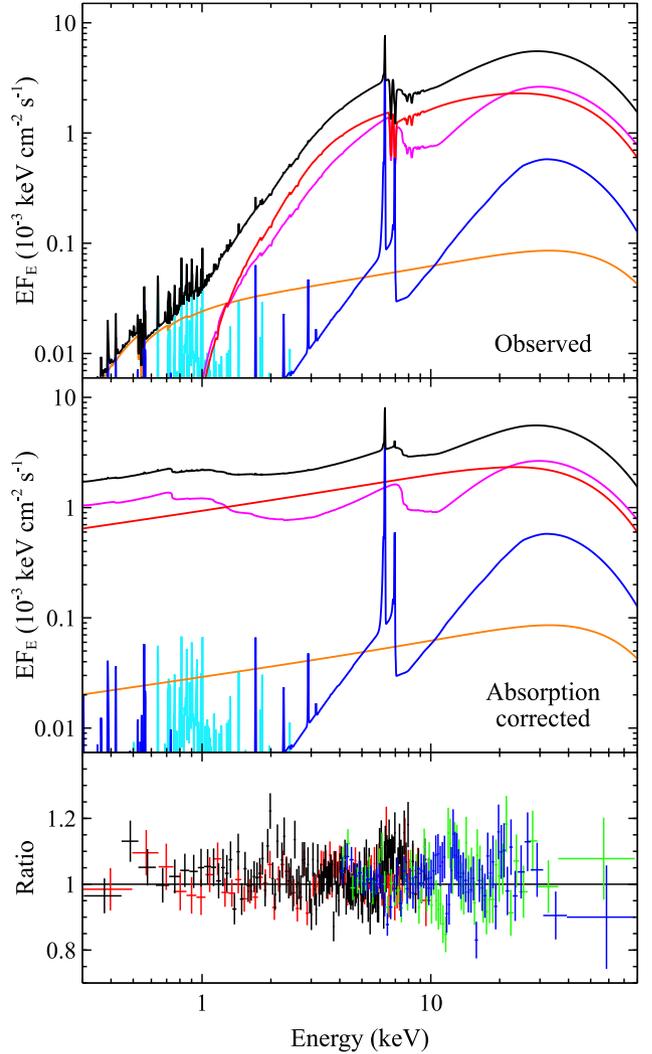}}
}
\end{center}
\vspace*{-0.3cm}
\caption{The relative contributions of the different components for our spectral model
for the coordinated \xmm+\nustar\ observation of \eso, using model 1a for illustration.
The total model is shown in black, the primary Comptonised continuum in red, the
relativistic disc reflection in magenta, the distant reflection in blue, the scattered
emission in orange and the Mekal plasma in cyan. We show the model both with (top
panel) and without (middle panel) the absorption components included. \textit{Bottom
panel:} The data/model ratio for this fit. As before, the data have been rebinned for
visual purposes, and the colours have the same meanings as in Figure \ref{fig_spec}.
}
\label{fig_model}
\end{figure}

\begin{table*}
  \caption{Parameter constraints for the various lamppost disc reflection models fit to
  the broadband \xmm+\nustar\ data available for \eso.}
\begin{center}
\begin{tabular}{c c c c c c c}
\hline
\hline
\\[-0.2cm]
Model Component & \multicolumn{2}{c}{Parameter} & \multicolumn{4}{c}{Model} \\
\\[-0.35cm]
& & & 1a & 1b & 2a & 2b \\
\\[-0.3cm]
\hline
\hline
\\[-0.1cm]
\tbabs$_{\rm{full}}$ & $N_{\rm{H}}$ & [$10^{22}$ cm$^{-2}$] & $1.2 \pm 0.2$ & $1.3 \pm 0.1$ & $1.2 \pm 0.1$ & $1.3 \pm 0.1$ \\
\\
\tbabs$_{\rm{part}}$ & $N_{\rm{H}}$ & [$10^{22}$ cm$^{-2}$] & $5.4^{+0.9}_{-0.7}$ & $5.9 \pm 0.5$ & $5.4 \pm 0.7$ & $5.8^{+0.3}_{-0.5}$ \\
\\[-0.3cm]
& $C_{\rm{f}}$ & [\%] &$79 \pm 2$ & $80^{+1}_{-2}$ & $79 \pm 2$ & $80^{+1}_{-2}$ \\
\\
\xstar\ & $\log\xi$ & $\log$[\ergcmps] & $3.45 \pm 0.05$ & $3.46^{+0.05}_{-0.04}$ & $3.46^{+0.04}_{-0.05}$ & $3.46^{+0.05}_{-0.04}$ \\
\\[-0.3cm]
& \nh\ & [$10^{22}$ cm$^{-2}$] & $5.1^{+2.6}_{-1.4}$ & $6.0^{+1.3}_{-1.0}$ & $5.3^{+2.0}_{-1.4}$ & $6.0^{+1.2}_{-1.0}$ \\
\\[-0.3cm]
& $v_{\rm{out}}$ & [\kmps] & $5400^{+600}_{-700}$ & $5400^{+600}_{-700}$ & $5400^{+600}_{-700}$ & $5400^{+600}_{-700}$ \\
\\
\relxilllpioncp/ & $\Gamma$ & & $1.70 \pm 0.07$ & $1.70 \pm 0.07$ & $1.71^{+0.06}_{-0.07}$ & $1.71^{+0.05}_{-0.07}$ \\ 
\\[-0.3cm]
\relxilllpdcp\ & $kT_{\rm{e}}$\tmark[a] & [keV] & $45^{+17}_{-8}$ & $46^{+17}_{-8}$ & $46^{+14}_{-8}$ & $50^{+20}_{-10}$ \\
\\[-0.3cm]
& $a^*$ & & $>0.96$ & $>0.96$ & $>0.96$ & $>0.96$ \\
\\[-0.3cm]
& $i$ & [\deg] & $51^{+5}_{-2}$ & $49^{+3}_{-1}$ & $50^{+3}_{-2}$ & $49 \pm 2$ \\
\\[-0.3cm]
& $h$ & [\rg] & $<2.1$ & $<2.0$ & $<2.1$ & $<2.0$  \\
\\[-0.3cm]
& \Rfrac\ & & $4.1^{+3.1}_{-1.0}$ & $5.3 \pm 1.5$\tmark[b] & $4.1^{+2.8}_{-1.0}$ & $5.3^{+1.7}_{-1.3}$\tmark[b] \\
\\[-0.3cm]
& $A_{\rm{Fe}}$\tmark[c] & [solar] & $4.9^{+1.8}_{-1.4}$ & $4.1^{+0.5}_{-0.8}$ & $4.8^{+1.7}_{-0.9}$ & $4.1^{+0.4}_{-0.5}$ \\
\\[-0.3cm]
& $\log\xi$ & $\log$[\ergcmps] & $3.1^{+0.5}_{-0.4}$ & $3.3^{+0.6}_{-0.2}$ & $3.1^{+0.3}_{-0.2}$ & $3.2 \pm 0.2$ \\
\\[-0.3cm]
& $p$ & & $<0.46$ & $<0.31$ & 0\tmark[d] & 0\tmark[d] \\
\\[-0.3cm]
& $\log n$ & $\log$[\pcmcub] & 15\tmark[d] & 15\tmark[d] & $<17.7$ & $<17.9$ \\
\\[-0.3cm]
& Norm & [$10^{-4}$] & $7.4 \pm 4.7$ & $5.5^{+1.7}_{-4.4}$ & $7.2^{+3.6}_{-3.9}$ & $5.8^{+4.6}_{-1.0}$ \\
\\
\xillvercp\ & Norm & [$10^{-6}$] & $7.8^{+2.3}_{-1.6}$ & $9.1^{+2.4}_{-1.7}$ & $7.9^{+2.3}_{-1.2}$ & $9.3^{+2.1}_{-1.7}$ \\
\\
\mekal\ & $kT$ & [keV] & $0.70^{+0.10}_{-0.06}$ & $0.70^{+0.09}_{-0.06}$ & $0.70^{+0.09}_{-0.06}$ & $0.70^{+0.09}_{-0.06}$ \\
\\[-0.3cm]
& Norm & [$10^{-6}$] & $7.7^{+2.0}_{-1.9}$ & $7.8^{+2.0}_{-1.8}$ & $7.8 \pm 1.8$ & $8.1 \pm 1.7$ \\
\\
\nthcomp$_{\rm{scat}}$ & Norm & [$10^{-5}$] & $2.9 \pm 0.4$ & $2.9^{+0.2}_{-0.4}$ & $2.9^{+0.3}_{-0.4}$ & $2.8^{+0.4}_{-0.3}$ \\
\\[-0.2cm]
\hline
\\[-0.25cm]
\chisq/DoF & & & 2509/2537 & 2510/2538 & 2509/2537 & 2510/2538 \\
\\[-0.25cm]
\hline
\hline
\end{tabular}
\label{tab_param}
\end{center}
\flushleft
$^a$ $kT_{\rm{e}}$ is quoted in the rest-frame of the X-ray source (\ie prior to any
gravitational redshift), based on the best-fit lamppost geometry. \\
$^b$ \Rfrac\ is calculated self-consistently for the lamppost geometry from
$a$ and $h$ for models 1b and 2b; the errors for these models represent the range
of values permitted by varying these parameters within their 90\% uncertainties. \\
$^c$ The iron abundance is linked across all spectral components associated
with the nucleus of \eso. \\
$^d$ These parameters are implicitly assumed in the \relxill\ variant used, and are not
free to vary.
\vspace{0.4cm}
\end{table*}

\subsection{Results}
\label{sec_fit}

The two variants of \relxill\ we consider in this work are \relxilllpioncp\ and \relxilllpdcp,
which we refer to as models 1 and 2, respectively. The former assumes a typical
density for reflection models available in the literature ($n = 10^{15}$\,\pcmcub) and
allows for an ionisation gradient across the disc, which we assume to have a
powerlaw form (\ie\ $\xi(r) \propto r^{-p}$, where $p$ is a free parameter; in this case
the ionisation parameter given is for the innermost disc). The latter assumes the disc
has a constant ionisation (\ie $p = 0$), but includes its density as a free
parameter.\footnote{Some aspects of this model are still in development, and as such
it is not distributed with current versions of \relxill\ as standard. In particular, the
additional high-density plasma effects discussed in \cite{Kallman21} are still being
incorporated. However, for the density regime we are interested in here ($n <
10^{19}$\,\pcmcub), the results should be reliable. The intention is to distribute
\relxilllpdcp\ with future versions of the \relxill\ package once these updates are
included.} We use these models because recent work in the literature has found that
both of these extensions to the standard reflection models typically used in the
literature (which typically assume either $p = 0$ or $n = 10^{15}$\,\pcmcub, or both)
may be important for modelling the X-ray spectra of AGN under certain circumstances
(\eg\ \citealt{Svoboda12, Kammoun19, Ingram19, Jiang18iras, Jiang19agn}).

\begin{figure}
\begin{center}
\hspace*{-0.35cm}
\rotatebox{0}{
{\includegraphics[width=235pt]{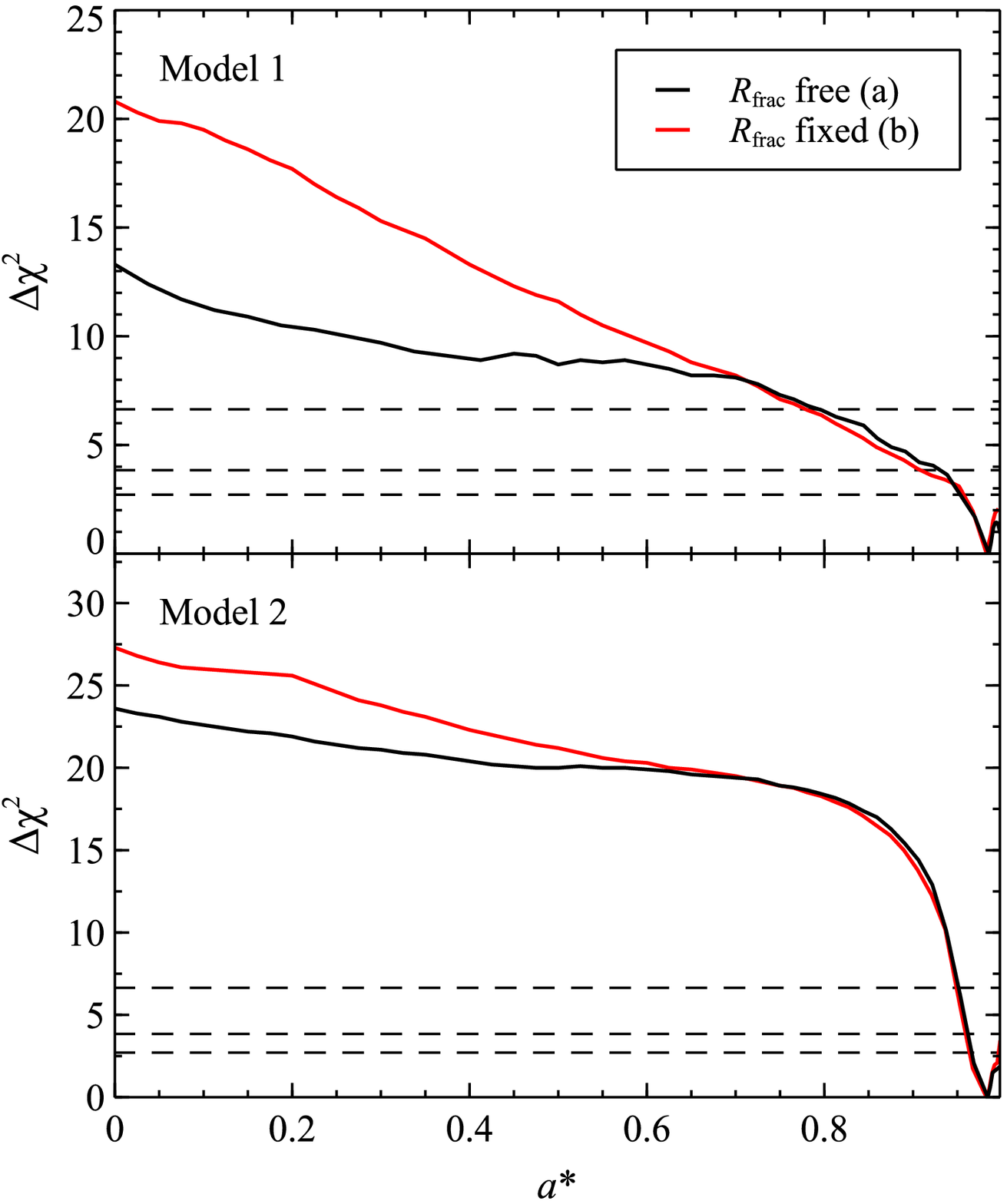}}
}
\end{center}
\vspace*{-0.3cm}
\caption{The $\Delta$\chisq\ confidence contours for the spin of \eso\ based on our
lamppost models for the coordinated \xmm+\nustar\ observation. Results for model 1
(\relxilllpioncp; variable ionisation gradient) and model 2 (\relxilllpdcp; variable disc
density) are shown in the top and bottom panels, respectively. In each case, we show
contours for the fits with \Rfrac\ free to vary (black) and computed self-consistently
from $a^*$ and $h$ in the lamppost geometry (red). The horizontal dotted lines show
the 90, 95 and 99\% confidence levels for a single parameter of interest.
}
\label{fig_spin}
\end{figure}

In both cases, we initially allow $R_{\rm{frac}}$ to vary as a free parameter (models
1a and 2a, respectively). Both of these models fit the data extremely well, with \chisq\
=  2509 for 2537 degrees of freedom (DoF), and both settle on essentially identical
best fits; the parameter constraints are given in Table \ref{tab_param}. For illustration,
we show the relative contributions of the various model components for model 1a in
Figure \ref{fig_model}, both with and without the inferred line-of-sight absorption,
along with the data/model ratio for this fit. Removal of any of the key model
components significantly degrades the fit, by $\Delta\chi^{2} > 20$ per degree of
freedom. We note in particular that removing the disc reflection degrades the fit by
$\Delta\chi^{2} \sim 300$ for 6 fewer free parameters. Despite their similarity, we still
present both models in order to explore how these different approaches influence
the constraints on the best-fit geometric parameters of interest (\ie $a^*$ and $h$).
The best fits imply a rather extreme scenario, involving a rapidly rotating black hole
($a^* > 0.95$) and an extremely compact X-ray source ($h \lesssim 2$\,\rg) in both
cases. We show the confidence contours for the spin constraints in Figure
\ref{fig_spin}. 

Such a combination would naturally predict a very large reflection fraction, as hinted
at by the previous data (\citealt{Balokovic17phd, Lanz19, Panagiotou19}), owing to
the strong gravitational lightbending associated with such a geometry. Indeed, the
best-fit reflection fractions are \Rfrac\ = 3--7 in both cases, significantly in excess of
the value expected for a standard accretion disc in the absence of lightbending
(which should give \Rfrac\ $\sim$ 1). This combination would also imply the
presence of a steep emissivity profile for the inner disc. Although we do not present
these fits in full, we therefore also explored a model in which the emissivity profile is
characterised by a broken powerlaw (\ie $\epsilon(r) \propto r^{-q_{\rm{in}}}$ for $R
< R_{\rm{br}}$ and $\propto r^{-q_{\rm{out}}}$ for $R > R_{\rm{br}}$) instead of a
formal lamppost profile. Fixing $q_{\rm{out}} = 3$ (\eg\ \citealt{Reynolds97feK}), we
find $q_{\rm{in}} > 5.3$ and $R_{\rm{br}} = 2.7^{+0.9}_{-0.2}$\,\rg, broadly
consistent with the profile expected for a very compact lamppost geometry (\eg\
\citealt{Wilkins11, Dauser13, Gonzalez17}). The other key reflection parameters
(\eg\ the spin and inclination constraints) are all consistent with those presented for
the lamppost models.

The strong disc reflection is primarily the reason that the exact choice of model for the
distant reflection does not matter here. The different geometries assumed in these
models result in differences in the relative strengths of the narrow iron emission and
the Compton hump from the distant reprocessor, and some differences in the exact
shape of the latter. While these differences can be important in some cases,
particularly in extremely obscured cases where the distant reflection  dominates the
spectrum, when the spectrum is dominated by reflection from the disc then these
differences have a negligible effect on the fit as the emission from the distant reflector
makes only a small contribution to the overall flux (see also  \citealt{Walton18}). As
noted above, this also means that the additional parameters relevant to \borus\ are
poorly constrained. We do not present the results in full for brevity, as they are
consistent with those already presented in Table \ref{tab_param}, but taking the
\relxilllpioncp\ model as an example, we find the covering factor of the torus to be
unconstrained above $\Omega/4\pi$ of 0.1 (the minimum value permitted by the
model), and the column density through the torus is only constrained to
$N_{\rm{H,tor}} > 2.5 \times 10^{22}$\,\pcmsq. Interestingly, this extends comfortably
into the Compton-thin regime, and is actually consistent with the line-of-sight column.
In that situation, there would be no Compton-hump associated with the distant
reprocessor at all, and \textit{all} of the reflection continuum would be associated with
the disc. With regards to estimating \Rfrac, adopting the \xillver\ model for the distant
reflector can therefore be considered a slightly conservative approach. Comparing
the relative fluxes of the \xillver\ and \relxill\ components in the 20--40\,keV band with
the \pexrav\ neutral reflection model (\citealt{pexrav}) implies a reflection fraction for
the distant reflector of $R_{\rm{frac,distant}} \sim 0.2$ with this model, within the
typical range seen from local Seyferts (\eg\ \citealt{Nandra07}).

We also find that the two extensions to standard reflection models considered here
are not formally needed by the data for \eso; with the \relxilllpioncp\ model the
ionisation gradient runs to zero, with an upper limit of $p < 0.46$, and with the
\relxilllpdcp\ model we find the density tends back to the typical value of
$10^{15}$\,\pcmcub, with an (fairly weak) upper limit of $\log[n/\rm{cm}^{-3}] < 17.7$.
This may be because of the lack of direct soft X-ray flux from the nucleus, which is
absorbed away below $\sim$1.5\,keV; although both $p$ and $n$ can have some
influence in the Fe K band, this soft band generally provides the majority of the
information that constrains these parameters. 

Given the model complexity, we also confirm the parameter constraints returned by
our \chisq\ analysis by running a series of Monte Carlo Markoff Chain (MCMC)
simulations, making use of the MCMC functionality within \xspec. We use the
Goodman-Weare algorithm (\citealt{MCMC_GW}), and focus on model 1a for this
exercise, taking the best-fit model parameters presented in Table \ref{tab_param} as
our starting point. The rest of the setup is: \nwalker\ walkers, each of which is run for 
\nstep\ steps with a burn-in length of \nburn, giving a total chain of \totchain\
parameter combinations. The full results of this run are presented in Appendix
\ref{app_mcmc}. Here, we simply note that the 90\% constraint on the spin found
from this analysis is $a^* > 0.97$, very similar to that obtained with our \chisq\
analysis.

\begin{figure}
\begin{center}
\hspace*{-0.35cm}
\rotatebox{0}{
{\includegraphics[width=235pt]{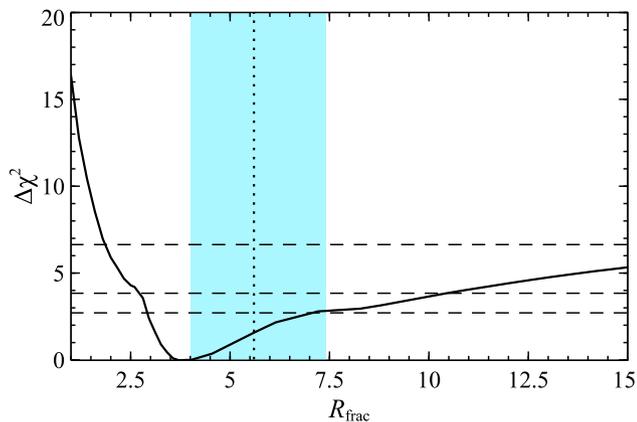}}
}
\end{center}
\vspace*{-0.3cm}
\caption{The $\Delta$\chisq\ confidence contour for \Rfrac\ when varied as a free
parameter in our analysis (solid black), focusing on Model 1 for illustration. The
horizontal dotted lines represent the same confidence levels as Figure
\ref{fig_spin}. The vertical dotted line shows the value of \Rfrac\ predicted by
$a^*$ and $h$ in the lamppost geometry, and the shaded region indicates the
range permitted by the 90\% statistical uncertainties on these parameters. There
is excellent correspondence between the observed and predicted values.
}
\label{fig_Rfrac}
\end{figure}

The best-fit reflection fractions found in both models 1a and 2a are quantitatively
consistent with the predicted reflection fractions based purely on $a^*$ and $h$ for
a thin disc model (predicted \Rfrac\ $\sim$ 5.4; see Figure \ref{fig_Rfrac}), broadly
similar to the recent case of IRAS\,09149--6206 (\citealt{Walton20iras}). We
therefore present further fits with each of the \relxill\ variants in which \Rfrac\ is
self-consistently computed from $a^*$ and $h$ in the lamppost geometry (models
1b and 2b). These are also presented in Table \ref{tab_param}; minimal differences
are seen in the best fits, but some of the parameter constraints have tightened
further (see below). The updated confidence contours for the spin are also shown
in Figure \ref{fig_spin}; although the formal 90\% constraints are not too different
for the spin specifically, these fits do exclude the low spin parameter space at
higher confidence than the fits in which \Rfrac\ is a free parameter.

In all of our models, the iron abundance is found to be significantly super-solar, with
best-fit abundances of $A_{\rm{Fe}} \sim$ 4--5. The constraints are tighter in models
1b and 2b, where we compute \Rfrac\ self consistently, as there is a mild degeneracy
between $A_{\rm{Fe}}$ and \Rfrac\ since both of these parameters play a role in
determining the strength of the iron emission line (see Appendix \ref{app_mcmc}). In
turn, this results in tighter constraints on the column densities of the various
absorbers, which are similarly connected to the iron abundance (also see Appendix
\ref{app_mcmc}). Interestingly, we find that the abundance constraints are similar for
both \relxill\ variants considered, implying the abundance does not depend on the
density here. Indeed, even if we force the density to the upper boundary of the range
for which \relxilllpdcp\ is currently valid ($n = 10^{19}$\,\pcmcub), we still find best-fit
abundances consistent with those reported for models 2a/b in Table \ref{tab_param}.

In other cases, allowing for higher densities can result in lower iron abundances being
inferred (\citealt{Tomsick18cyg, Garcia18, Jiang19agn}), as increasing the disc density
can also influence the reflection continuum in the iron band (in addition to the main
changes that occur at low energies), and in turn the best-fit iron abundance. For \eso\
specifically, the fact that this does not occur may be because the high iron abundance
is not only driven by the disc reflection. If we split the main spectral components into
those associated with the disc (the relativistic reflection and the ionised absorption)
and those presumably associated with more distant nuclear structures (the neutral
absorbers and the distant reflector), and allow the two groups to have different iron
abundances, using model 2b we still find a super-solar iron abundance of
$A_{\rm{Fe,distant}} = 3.6^{+2.6}_{-1.5}$ for the `distant' group, consistent with the
constraint found for the `disc' group of $A_{\rm{Fe,disc}} = 4.3^{+3.8}_{-0.8}$; a solar
abundance for the `distant' group is strongly disfavoured ($\Delta\chi^{2} = 17$).
\cite{Reynolds12} also suggested an alternative `radiative levitation' explanation for
the super-solar iron abundances seen in some AGN, in which radiation pressure
within the disc preferentially forces iron to the surface, artificially enhancing the
perceived abundance. However, again the consistently high abundance seen for both
the disc and the distant model components would suggest this is not relevant here,
and that \eso\ really does have a super-solar iron abundance.

\section{Discussion}
\label{sec_dis}

We have presented a detailed analysis of the the first broadband (\ie\ \xmm+\nustar)
observation of the type-2 Seyfert \eso. These data show clear evidence for X-ray
obscuration, broadly consistent with its type-2 optical classification and qualitatively
similar to the limited archival observations available for this source (\citealt{Smith96,
Vignali98, Marchesi17}, although note that the final absorption model we use is more
complex than those considered in these works); the direct emission from the nucleus
is completely absorbed away below $\sim$1.5\,keV, leaving only scattered/diffuse
emission at the lowest energies, and there is clear evidence for reprocessing by
distant material in the form of a narrow emission line from neutral iron. However, the
level of obscuration is not so strong as to prevent us from viewing the innermost
regions of the accretion flow: a strong contribution from relativistic disc reflection is
also seen at higher energies via a relativistically broadened iron emission line and its
associated Compton reflection continuum, and an ionised outflow is also seen via
blueshifted absorption from \fexxv\ and \fexxvi. The combined coverage from \xmm\
and \nustar\ allows us to disentangle all of these contributions to the observed
spectrum, and to place robust constraints on the emission from the nuclear regions
despite their moderate obscuration.

\subsection{X-ray Constraints on the Black Hole}

By modelling the reflection from the innermost accretion disc, we are able to place
tight constraints on the spin of the black hole. Although we have used two different
reflection models (allowing for a radial ionisation gradient across the disc and for a
variable density), and made differing assumptions about the precise treatment of the
reflected emission in our analysis (allowing the reflection fraction to be both coupled
and decoupled to the spin constraints), we find consistent results for the spin across
all of our models: $a^* > 0.96$ (at 90\% confidence). Systematic errors on spin
measurements in this regime are likely $\lesssim$ 0.1 (\eg\ \citealt{Bonson16,
Choudhury17, Kammoun18}). \eso\ therefore joins the growing list of rapidly rotating
SMBHs in active galactic nuclei (\eg\ \citealt{Risaliti13nat, Walton13spin, Walton14,
Walton20iras, Parker14mrk, Reis14nat, Reynolds14, Chamani20}; see
\citealt{Reynolds20rev} for a recent review). Similar to many of the other AGN with
rapidly rotating black holes, \eso\ is radio-quiet, arguing further against the idea that
the difference between radio-loud and radio-quiet AGN is solely related to different
black hole spins, with high/low spin black holes exhibiting/lacking jet activity (as has
been suggested previously, \eg\ \citealt{Wilson95, Moderski98, Sikora07}). We stress,
however, that this only means that black hole spin is not simply a `switch' that
determines whether a jet is launched or not; when jets \textit{are} launched, it is
certainly still plausible that the spin could play an important role in setting the jet
power (\eg\ \citealt{BZ77}).

The spin of a SMBH is dictated by the manner in which its most recent major growth
episode occurred. Growth by a series of chaotic accretion events/mergers will tend
to spin the black hole down, while growth via prolonged `coherent' accretion (in
which the accreted material has a common angular momentum axis) will spin the
black hole up (\eg\ \citealt{King06, Sesana14, Fiacconi18}). The high spin of the
SMBH in \eso\ suggests its most recent growth episode is best described with the
latter scenario. Indeed, the barred-spiral morphology of the host galaxy shows no
obvious evidence for disruption related to recent merger activity (\citealt{Malkan98})
which could have triggered a more chaotic recent accretion history.

Ultimately, though, in terms of informing black hole growth/galaxy evolution models
on a global scale, it is the distribution of SMBH spins that is key (\eg\ \citealt{Berti08,
Sesana14}). Some preliminary attempts have been made to build this distribution in
the literature (\citealt{Walton13spin, Vasudevan16}), which show a preference for
higher spins, but these have been hampered by the relatively low numbers of
reflection-based measurements available (even today, we only have $\sim$35 AGN
spin constraints; \citealt{Reynolds20rev}) and the fact that there is a known
observational bias towards rapidly rotating black holes (as such sources should be
brighter for a given accretion rate; \citealt{Brenneman11}). In order to overcome
these limitations, the sample of AGN with spin constraints still needs expanding. In
this respect, the impact of \nustar\ will continue to be significant, as its broadband
coverage this has opened up the possibility of obtaining robust spin measurements
for even slightly more obscured systems by accurately determining the contribution
from reflection (\eg\ \citealt{Risaliti13nat, Walton14, Walton18, Walton19ufo}). In
these cases, such constraints would have been challenging with only soft X-ray
coverage owing to modelling degeneracies, and we can now add \eso\ to this list.
Although numbers are still small, early indications are that these more obscured
cases also show a preference for rapidly rotating black holes, similar to their less
obscured counterparts, but further measurements are required here too before
robust conclusions can be made.

The mass of the black hole in \eso\ is not currently well established in the literature;
as a type-2 Seyfert, there are no optical broad lines from which to determine the
mass via either the line widths or via optical reverberation mapping campaigns. 
Nevertheless, we can obtain a very rough estimate from its X-ray properties. The
X-ray photon index is known to correlate with the Eddington ratio of the black hole,
$\lambda_{\rm{E}} \equiv L_{\rm{bol}} /  L_{\rm{Edd}}$ (\eg\ \citealt{Shemmer08, 
Risaliti09LxGam, Brightman13, Trakhtenbrot17}). The photon index of $\Gamma
\sim 1.7$ found here would imply $\lambda_{\rm{E}} \sim 0.02$, based on the more
recent of these works. We can also estimate $L_{\rm{bol}}$ from the 2--10\,keV
luminosity inferred from our work via the appropriate bolometric correction
$\kappa_{2-10} \equiv L_{\rm{bol}} / L_{2-10}$; this correction is known to be a
function of Eddington ratio (\citealt{Vasudevan09, Lusso10}), and $\lambda_{\rm{E}}
\sim 0.02$ would imply $\kappa_{2-10} \sim 10$. From our spectral fits we find the
observed and absorption-corrected 2--10\,keV fluxes to be $F_{\rm{2-10,obs}} \sim
4.3 \times 10^{-12}$\,\ergpcmsqps\ and $F_{\rm{2-10,deabs}} \sim 6.7 \times
10^{-12}$\,\ergpcmsqps, respectively. The latter corresponds to an intrinsic
2--10\,keV luminosity of $L_{2-10} \sim 5 \times 10^{42}$\,\ergps\ for a luminosity
distance of 77\,Mpc.\footnote{Based on $z = 0.0181$ and assuming a standard
$\Lambda$CDM concordance cosmology, \ie\ \lambdaCDM.} Combining this with
$\lambda_{\rm{E}} \sim 0.02$ and $\kappa_{2-10} \sim 10$ gives $L_{\rm{bol}} \sim
5 \times 10^{43}$\,\ergps\ and $\log[M_{\rm{BH}} / \msun] \sim 7.3$. However, given
both the statistical uncertainty on $\Gamma$ and the scatter on the $\kappa_{2-10}$
vs $\lambda_{\rm{E}}$ trend, the uncertainty on this estimate is likely almost an order
of magnitude.

We can also estimate the mass from the X-ray variability observed, since the lack of
strong spectral variability suggest this is largely intrinsic to the AGN (as opposed to
being driven by changes in the line-of-sight absorption; Figure \ref{fig_lc}). Although
we do not have sufficient data for more sophisticated methods, such as using
features in the variability power spectrum (\eg\ \citealt{McHardy06}), we use the
anti-correlation between the normalised excess variance ($\sigma_{\rm{NXS}}$, see
\citealt{Vaughan03}) and black hole mass for local AGN reported by \cite{Ponti12var}.
We follow their approach and calculate $\sigma_{\rm{NXS}}$ on 40\,ks timescales
after having rebinned the 2--10\,keV \xmm\ lightcurve to 250s bins, and find
$\sigma_{\rm{NXS}} = 0.011 \pm 0.002$. This would imply $\log[M_{\rm{BH}} /
\msun] \sim 7.0$ based on the trend reported for AGN with reverberation-mapped
masses, in good agreement with the mass estimated from the spectral results, but
again with fairly significant uncertainties (roughly a factor of $\sim$5, combining the
scatter reported in \citealt{Ponti12var} and the systematic error on reverberation
mapped masses of $\sim$0.4 dex discussed by \citealt{Peterson14rev}).

Although both of the mass estimates from the X-ray spectroscopy and the X-ray
timing for \eso\ carry significant uncertainties, they are in encouraging agreement.
Furthermore, these X-ray mass estimates are also in good agreement with the mass
estimate of $\log[M_{\rm{BH}} / \msun] = 7.5 \pm 0.4$ recently obtained by the BAT
AGN Spectroscopic Survey team (BASS; \citealt{Koss17}) based on VLT/Xshooter
optical spectroscopy. These data allow a measurement of the stellar velocity
dispersion of the host galaxy, from which the back hole mass is inferred based on
standard scaling relations (\eg\ \citealt{Kormendy13rev}). Appendix \ref{app_mass}
briefly summarizes these optical data and their analyiss; full details will be presented
in Koss et al. (\textit{in prep.}). Improved mass constraints may be possible in the
future via reverberation mapping of the relativistic iron line (\eg\ \citealt{Alston20iras}),
but such work would require a series of long X-ray observations.

\subsection{The X-ray Source}

Although the X-ray corona is widely expected to be related to Compton up-scattering
of thermal photons emitting by the accretion disc (\eg\ \citealt{Haardt91}), the
detailed properties of this structure (its geometry, temperature, etc.) are still being
explored. Through a variety of arguments, combining intrinsic X-ray
variability/reverberation (\eg\ \citealt{FabZog09, Zoghbi12, Kara16}), microlensing
variations in lensed quasars (\eg\ \citealt{Chartas09, Dai10, MacLeod15}), X-ray
occultation events (\eg\ \citealt{Risaliti09a, Gallo21}) and detailed analyses of the 
emissivity profiles implied by relativistic disc reflection models (\eg\ \citealt{Wilkins11,
Fabian13iras}), we know the corona must typically be a compact region close to the
black hole (within $\sim$10\,\rg\ or so). However, the strength of the disc reflection
found in our analysis (\Rfrac\ $\sim$ 5) implies a particularly compact corona in \eso;
assuming a lamppost geometry with a thin disc, we find that the corona must be
within $\sim$2\,\rg\ of the black hole in order to produce the necessary lightbending.
Similar reflection-dominated states have been seen in other AGN (\eg\
\citealt{Fabian12, Parker14mrk}), where coronae within a few \rg\ of the black hole
have also been inferred.

Although a lamppost is clearly a simplified geometry (\citealt{Wilkins12, Zhang19}),
the requirement for such a compact corona reinforces the use of such models in our
spectral analyses; the more compact the corona, the more reasonably it can be
approximated as a lamppost. Disc geometries beyond the razor-thin approximation
used in \relxill\ have recently been explored in the literature. In particular, \cite{fenrir}
allow for a varying scale-height for the accretion disc, following the expectation that
at higher accretion rates larger scale-height discs are expected (\citealt{Shakura73}).
If the scale-height of the disc is non-negligible then the inner regions can form more
of a funnel-like geometry, which could potentially help to enhance the reflection
fraction. However, this would only have a significant effect if the size of the X-ray
source is smaller than the scale-height of the funnel, and in most cases this would
imply that the corona is likely to be in a regime in which strong gravitational
lightbending is important anyway. Furthermore, the Eddington ratio inferred above
is very modest, $\lambda_{\rm{E}} \sim 0.02$, and the disc is still expected to be
geometrically thin in this regime as radiation pressure should not play a major role,
so it is not clear that such a thick disc geometry is a plausible alternative explanation
for the large \Rfrac\ found in this case. Furthermore, we note that the constraints on
$h$ are still tight even in the models where this is decoupled from \Rfrac. We
therefore conclude that the X-ray corona really is very compact in this system.

Despite this, we note that it should still be possible to accommodate a corona that 
intercepts sufficient disc photons to power the observed X-ray flux within the relevant
size scales for \eso. Following the approach of \citet[][see their equation
6]{Dovciak16} we estimate the minimum `size' for the corona ($dh$) required by the 
assumption that the X-rays are powered by inverse Compton scattering of disc
photons. The type-2 nature of \eso\ means that the thermal emission from the disc
cannot be observed directly in this case, but adopting the above X-ray bolometric 
correction of $\kappa_{2-10} \sim 10$ as being representative of the relative X-ray
and disc fluxes, we find that the key results here ($\Gamma$, $kT_{\rm{e}}$, $a^*$
and $h$) comfortably satisfy $dh < h$, as the minimum size implied is only $dh \sim
0.2$\,\rg. This is broadly consistent with the results of \cite{Ursini20}, who find that
solutions with $dh < h$ are generally available for rapidly rotating black holes.

Taken at face value, the fact that strong reflection has been reported for both
\nustar\ observations, taken in 2014 and 2020 could imply that black hole coronae
can remain highly compact for extended periods of time ($\sim$several years).
However, we cannot rule out the possibility that the source geometry is variable on
these timescales either. Indeed, extrapolating beyond the \nustar\ bandpass, the
14--195\,keV flux implied by our model is $F_{14-195} \sim 10^{-11}$\,\ergpcmsqps,
a factor of $\sim$2.5 lower than the long-term average seen by \swift/BAT
($F_{14-195} = 2.5^{+0.2}_{-0.3} \times 10^{-11}$\,\ergpcmsqps; \citealt{BAT105m}).
This may imply we caught \eso\ in an unusually low-flux state, and in turn imply a
variable coronal geometry (\eg\ \citealt{Parker14mrk, Fabian12}), but further deep
\nustar\ observations will be required to confirm this.

In addition to the geometric constraints from the disc reflection, the high-energy
coverage from \nustar\ also provides insight into the plasma conditions in the corona,
and in particular the temperature of the Comptonising electrons (\kte) which governs
the high-energy cutoff in the primary continuum emission (\eg\ \citealt{Brenneman14a,
Matt15, Balokovic15, Garcia15, Walton16cyg, Kamraj18}). As noted above, the
variants of the \relxill\ model we use here both assume the primary continuum is a
thermal Comptonisation model, and so provide constraints on \kte\ directly, and our
use of the lamppost geometry variants means that this is evaluated in the rest-frame
of the X-ray source (\ie\ gravitational redshift is accounted for; note that this produces
the mild dependence seen between \kte\ and the spin in the Monte Carlo contours, as
the spin helps set $z_{\rm{gra}}$). Beyond these assumptions, the results are largely
independent of the precise details of the reflection model used, and we find \kte\ =
40--70\,keV. 

This differs from the initial estimate based on a shorter \nustar+\swift\ snapshot of
$E_{\rm{cut}} > 480$\,keV (when modelling the high-energy cutoff as an exponential;
\citealt{Balokovic20}); adopting a fairly standard conversion of $E_{\rm{cut}} =
3kT_{\rm{e}}$ (\citealt{Petrucci01}), our measurement would correspond to
$E_{\rm{cut}}$ = 120-210\,keV (note also that the previous constraint is in the
observed frame, and does not consider any gravitational redshift, which would further
increase the discrepancy). However, as noted by \cite{Balokovic20}, their result
marked \eso\ as something of an outlier among the broader AGN population. The
electron temperature obtained here is much more in line with typical results for other
AGN (\eg\ \citealt{Fabian15, Tortosa18, Ricci18, Balokovic20}), demonstrating the
importance of both high S/N broadband coverage and detailed modelling of the
reflection and absorption in these complex cases when determining \kte\ as well as
geometric constraints on the X-ray source. This is not just an issue relating to lower
S/N giving poorer statistical constraints on model parameters; the low S/N also
forces the use of only simple models. Although such models can still be informative
in many cases, for complex sources such as \eso\ they can give misleading results.

\cite{Fabian15} compiled the constraints from accreting black holes (both AGN and
X-ray binaries) for which \kte\ has been constrained by high S/N \nustar\ data,
characterising the data in terms of the compactness--temperature plane (here
`compactness' refers to the radiative compactness; this can be expressed as $l =
4\pi m_{\rm{p}} \lambda_{\rm{E}} / m_{\rm{e}} R_{\rm{X}}$, where $R_{\rm{X}}$ is
the size of the X-ray source in \rg, and $m_{\rm{e}}$ and $m_{\rm{p}}$ are the
electron and proton masses, respectively). In doing so, \cite{Fabian15} were able to
show that the results cluster close to the limit set by pair runaway, the point at which
further injection of energy into the corona no longer increases its temperature but
instead results in increasing pair generation (see also \citealt{Fabian17} and
\citealt{Tortosa18} for updated samples of \kte\ constraints with \nustar). The
properties of X-ray coronae therefore appear to be primarily dictated by this
pair-production `thermostat'. Given that we infer the corona to be highly compact (in
a geometrical sense) here, we also consider where \eso\ lies in the
$l$--$kT_{\rm{e}}$ plane. Taking the size of the X-ray source to be $R_{\rm{X}} \sim
h$, and $\lambda_{\rm{E}} \sim 0.02$, we therefore find $l \sim 230$. Combining
this with our temperature constraints implies that the X-ray source in \eso\ is also
consistent with being close to this pair runaway limit. \eso\ therefore does not
appear to be abnormal in this regard, despite its highly compact corona.

\subsection{The Low-Energy Absorption and the Ionised Outflow}

Even though the covering factor is fairly high ($\sim$80\%) the primary low-energy
absorber (in terms of column density) does not appear to cover the whole X-ray
source. Geometrically this would be quite challenging to achieve for a very distant
absorber, given how compact the X-ray source appears to be. We therefore
speculate that this component may be similar to the transient `obscurer' seen
recently in NGC\,5548 (\citealt{Kaastra14}). This low-ionisation
($\log[\xi/(\rm{erg}~\rm{cm}~\rm{s}^{-1})] \sim -1.2$) outflow persisted for several
years, covered a similar fraction of the X-ray source ($\sim$85\%), and was
inferred to be a stream of material originally launched from the accretion disc that
also extended sufficiently to obscure some fraction of the broad line region (BLR),
and so its location was deemed to be comparable to the latter. Several similar
events have now been recognised in other AGN (\eg\ \citealt{Ebrero16,
Mehdipour17, Gallo21}). Allowing the partially covering component to be
outflowing does not offer any notable improvement to the fit here, but we find the
upper limit on any outflow velocity to be $\sim$5000\,\kmps, which would still
permit a similar outflow velocity to that seen from the NGC\,5548 obscurer. The
X-ray source is likely more compact by a factor of a few in the case of \eso\ (the
X-ray source size was assumed to be $\sim$10\,\rg\ in the case of NGC\,5548),
which would in turn require the absorber to be closer by a similar factor in
comparison (the black hole masses are quite similar), but the overall comparison
seems promising. This could imply that the partially covering component could be
transient for \eso\ as well, and it is therefore interesting to note that the total level
of obscuration inferred here seems to exceed that of the archival observations, for
which $N_{\rm{H}} \sim 1-2 \times 10^{22}$\,\pcmsq\ (similar to our fully-covering
component; \eg\ \citealt{Smith96, Balokovic17phd}). Further monitoring of \eso\
with high S/N broadband observations would therefore be of significant interest.

The ionised absorption seen at $\sim$7\,keV is clearly blueshifted relative to the
cosmological redshift of \eso, implying an outflow with a line-of-sight velocity of
$\sim$5400\,\kmps. Although this is fast relative to typical outflow velocities seen in
lower ionisation warm absorbers (typically $v_{\rm{out}} \sim 1000$\,\kmps, \eg\
\citealt{Laha14}), similar velocities are seen for highly ionised outflows in a number
of other systems (\eg\ \citealt{Risaliti05b, Walton18, Walton20iras}), and the wind in
\eso\ does not come close to the most extreme winds seen from AGN in this respect
(the `ultrafast' ouflows, which can reach $v_{\rm{out}} \sim 0.4c$, \citealt{Reeves18,
Walton19ufo}). Following \cite{Tombesi12}, we attempt to constrain the location of
the wind by considering both escape velocity and ionisation arguments; the former
indicates the lower bound on the wind radius ($R_{\rm{w}}$, assuming the wind
escapes, \ie $R_{\rm{w}} > R_{\rm{esc}}$, where $R_{\rm{esc}}$ is determined by
assuming $v_{\rm{esc}} = v_{\rm{out}}$ and the mass determined above), while the
latter likely provides an upper bound (see also \citealt{Nardini15}). Here, we find
$10^{16} \lesssim R_{\rm{w}} \lesssim 10^{17}$\,cm (or equivalently $5,000
\lesssim R_{\rm{w}} \lesssim 50,000$\,\rg). This is somewhat intermediate to the
range of radii inferred for the ultrafast outflows and the classic warm absorbers
(\citealt{Tombesi12}), and in the case of \eso\ likely places the outflow in the BLR.
We also attempt to estimate the kinetic power carried by the wind in \eso\ relative
to its radiative output ($L_{\rm{kin}}/L_{\rm{bol}}$) following the same methodology
as \cite{Walton20iras}, \ie combining equation 1 from that work (in which
$L_{\rm{kin}} / L_{\rm{bol}}$ is expressed in terms of the column density, outflow
velocity and the radius of the wind) with the assumption that $R_{\rm{w}} >
R_{\rm{esc}}$, since this may provide the lower bound on $R_{\rm{w}}$. With this
approach, we find $L_{\rm{kin}}/L_{\rm{bol}} > 0.04 ~ \Omega C_{\rm{V}}$ (where
$\Omega$ and $C_{\rm{V}}$ are the solid angle subtended by the wind and its
volume filling factor, respectively, both of which are  normalised to the range 0--1).

The observed co-evolution between SMBHs and their host galaxies (\eg\
\citealt{Ferrarese06, Kormendy13rev}) requires some kind of feedback mechanism
that is able to connect their disparate size scales, and accretion-driven outflows are
considered a prime candidate for this (see \citealt{Fabian12rev} for a review). In
order for this to be the case, simulations suggest that $L_{\rm{kin}}$ must be at
least a few per cent of \lbol\ (\eg\ \citealt{DiMatteo05nat, Hopkins10}). Although the
outflow velocity is fairly modest in this case, the relative kinetic power of the wind in
\eso\ still appears to be fairly close to this threshold. It should be noted that owing
to our definitions, the product $\Omega C_{\rm{V}}$ must be $\leq 1$. Furthermore,
we note again that significant uncertainties remain on the mass, and also likely on
$L_{\rm{bol}}$, both of which translate into additional significant uncertainties on
$L_{\rm{kin}}/L_{\rm{bol}}$. Depending on how these factors balance with the size
of the launch radius (relative to $R_{\rm{esc}}$), the outflow in \eso\ may plausibly
carry sufficient energy to drive galaxy-scale feedback, but a firm case for this
conclusion cannot yet be made.

\section{Conclusions}
\label{sec_conc}

The nuclear X-ray spectrum of the type-2 Seyfert \eso\ is complex, showing evidence
for absorption by both neutral and ionised material, as well as reprocessed emission
from both the accretion disc and more distant material. Utilizing the broadband
coverage provided by \xmm\ and \nustar, we have been able to disentangle all of
these various contributions to the observed data. The total neutral absorption column
is fairly large, $N_{\rm{H}} \sim 5-6 \times 10^{22}$\,\pcmsq, consistent with its optical
classification as a type-2 Seyfert but not so large as to prevent us from viewing the
innermost accretion flow, while the ionised absorption occurs in a moderately rapid
outflow ($v_{\rm{out}} = 5400^{+600}_{-700}$\,\kmps) with a column comparable to
the neutral absorption. Most remarkably, the contribution from reflection from the
innermost accretion flow is very large, \Rfrac\ $\sim$ 5, requiring strong gravitational
lightbending. In turn, this requires both an extremely compact X-ray corona and a
rapidly rotating black hole; we find that the corona must be within $\sim$2\,\rg\ of the
black hole, which must have a spin of $a^* > 0.96$. \eso\ is therefore yet another
addition to the sample of radio-quiet AGN that host high-spin black holes. In addition
to its geometry, the high-energy coverage from \nustar\ also allows us to place
constraints on the temperature of the corona, which we find to be $kT_{\rm{e}} =
40-70$\,keV. Although its geometry is fairly extreme, similarly compact coronae have
been inferred in a handful of other AGN, and when considered in terms of the
compactness--temperature plane, \eso\ does not seem abnormal when compared
with the broader AGN population, lying similarly close to the limit set by runaway pair
production.

\section*{ACKNOWLEDGEMENTS}

%The authors would like to thank the reviewer for their positive feedback, which
%helped to improve the final version of the manuscript.
DJW acknowledges support from the Science and Technology Facilities Council
(STFC) in the form of an Ernest Rutherford Fellowship (ST/N004027/1).
MB acknowledges support from the YCAA Prize Postdoctoral Fellowship.
CSR thanks the STFC for support under Consolidated Grant ST/S000623/1, as well
as the European Research Council (ERC) for support under the European Union’s 
Horizon 2020 research and innovation programme (grant 834203).
CR acknowledges support from the Fondecyt Iniciacion grant 11190831.
JAG acknowledges support from NASA grant 80NSSC19K1020  and from the
Alexander von Humboldt Foundation.
This research has made use of data obtained with \nustar, a project led by Caltech,
funded by NASA and managed by the NASA Jet Propulsion Laboratory (JPL), and
has utilized the \nustardas\ software package, jointly developed by the Space
Science Data Centre (SSDC; Italy) and Caltech (USA).
This research has also made use of data obtained with \xmm, an ESA science
mission with instruments and contributions directly funded by ESA Member States.
This work has made use of the \textsc{corner} package (\citealt{corner}) for data 
visualisation.

%{\it Facilities:} \facility{NuSTAR}, \facility{XMM-Newton}, \facility{Chandra}

\section*{Data Availability}

All of the data underlying this article will be available from June 2021 via ESA's
\xmm\ Science Archive (https://www.cosmos.esa.int/web/xmm-newton/xsa) and
NASA's HEASARC archive (https://heasarc.gsfc.nasa.gov/).

\appendix

\section{Monte Carlo Results}
\label{app_mcmc}

\begin{figure*}
\begin{center}
\hspace*{-0.1cm}
\rotatebox{0}{
{\includegraphics[width=470pt]{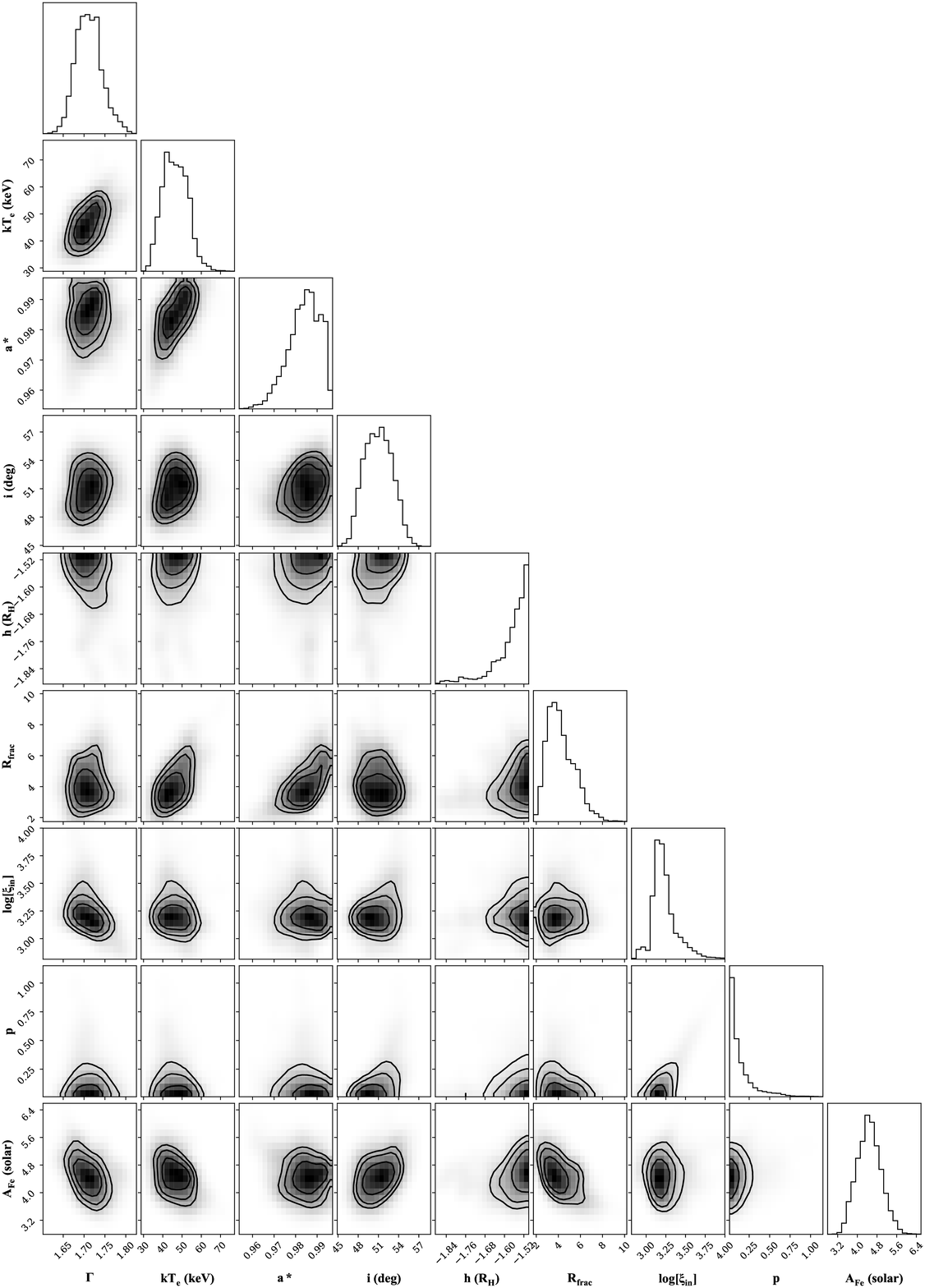}}
}
\hspace{0.6cm}
\end{center}
\vspace*{-0.3cm}
\caption{The MCMC results for parameters relating to the intrinsic AGN continuum
and the relativistic disc reflection for model 1a. The 2-D contours show the 1, 2 and
3$\sigma$ confidence levels based on the 2-D MCMC parameter densities.
}
\label{fig_app_mcmc1}
\end{figure*}

\begin{figure*}
\begin{center}
\hspace*{-0.1cm}
\rotatebox{0}{
{\includegraphics[width=460pt]{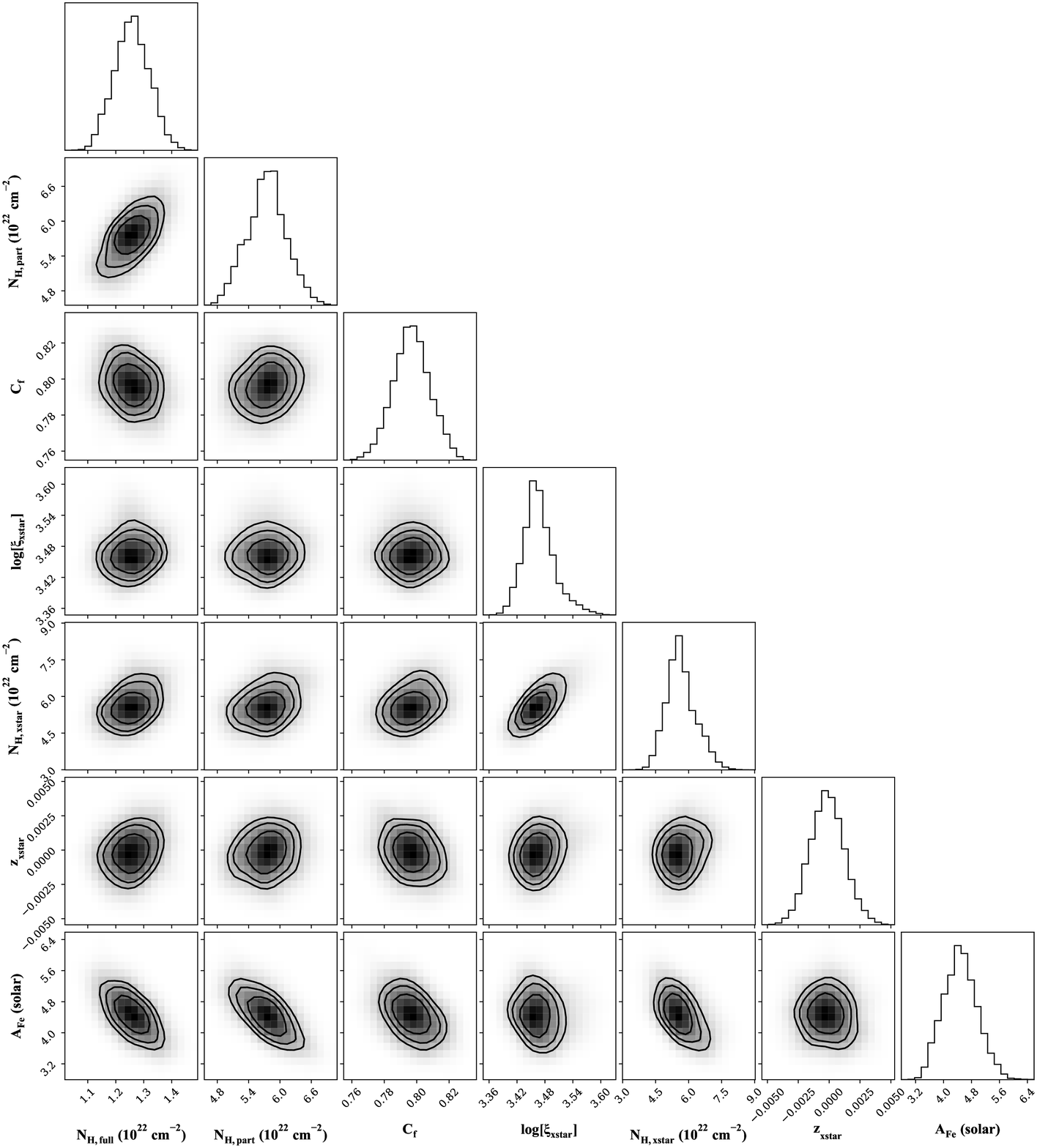}}
}
\hspace{0.6cm}
\end{center}
\vspace*{-0.3cm}
\caption{The MCMC results for parameters relating to the various absorbers in
model 1a. The plot format follows that of Figure \ref{fig_app_mcmc1}.
}
\label{fig_app_mcmc2}
\end{figure*}

Here we present the results from our MCMC simulations performed for model 1a
(see Section \ref{sec_fit}). For clarity, we focus on the parameters relating to the
intrinsic continuum and the disc reflection in Figure \ref{fig_app_mcmc1}, and on the
parameters relating to the various absorbers in Figure \ref{fig_app_mcmc2}. Note
that here, the lamppost height is in units of the vertical horizon (\rh) which varies
from $1 \leq R_{\rm{H}} / R_{\rm{G}} \leq 2$, depending on the spin (the negative
values are related to the \relxill\ setup). In addition, the outflow velocity of the ionised
absorber is given in terms of its redshift in the observed frame. We also stress that
although the iron abundance is shown in both Figure \ref{fig_app_mcmc1} and
\ref{fig_app_mcmc2} as it is important to both the reflection and the absorption, the
abundance is common to all of these components; showing it in both also allows the
reader to draw links between the panels in the two Figures. Chain convergence is
good for all model parameters, with the convergence measure proposed by
\cite{Geweke92} always close to zero.

\begin{figure*}
\begin{center}
%\hspace*{-1.0cm}
\rotatebox{0}{
{\includegraphics[width=475pt]{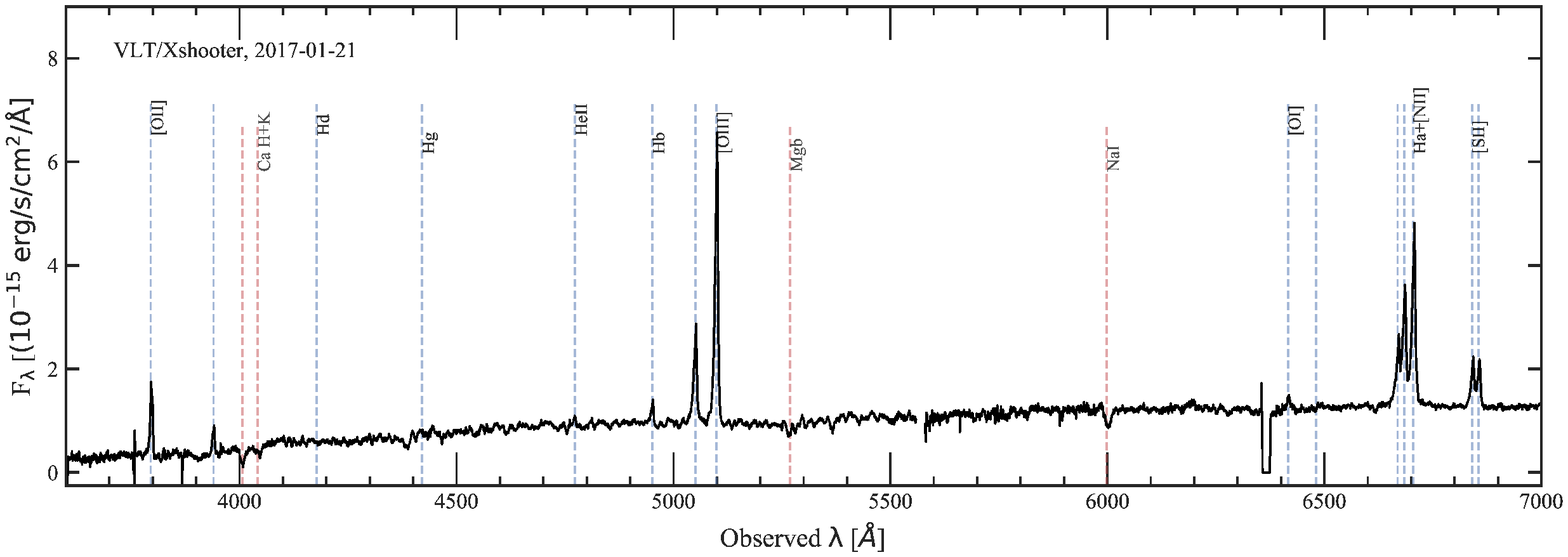}}
}
\hspace{0.6cm}
\end{center}
\vspace*{-0.2cm}
\caption{VLT/Xshooter data for \eso\ (see Koss et al. \textit{in prep.} for full details). In
addition to the strong, narrow emission lines, typical for a Seyfert 2 galaxy, stellar
absorption features can also be seen, allowing for an estimate of the stellar velocity
dispersion, and in turn the black hole mass (see text). Line identifications are indicated, 
with emission lines shown in cyan and absorption features shown in light brown.
}
\label{fig_app_opt}
\end{figure*}

\section{Optical Constraints on the Black Hole Mass}
\label{app_mass}

As part of ongoing efforts to follow-up AGN detected in hard X-rays by \swift/BAT,
the BASS survey team\footnote{https://www.bass-survey.com/} recently obtained
updated optical spectroscopy for \eso\ with VLT/Xshooter (\citealt{XSHOOTER}).
These data revealed resolved stellar absorption features, providing a measurement
of the central stellar velocity dispersion, $\sigma_{*}$. In turn, this provides an
independent estimate of the black hole mass using the well-known $M_{\rm{BH}} -
\sigma_{*}$ relation (\citealt{Kormendy13rev}). Full details of the spectroscopy and
analysis performed will be presented in Koss et al (\textit{in  prep.}), here we
summarise a few of the salient details relevant to this optical mass constraint.

\eso\ was observed for a 480\,s exposure on UT 2017-01-21, with the Xshooter
data relevant here coming from the UVB and VIS arms (3000--5595 and
5595--10,240\,\AA, respectively). The wavelength ranges 3880-5550\,\AA\ and
8350-8730\,\AA\ were used specifically to constrain $\sigma_{*}$, as these cover key
stellar absorption features from \Cahk\ (3935, 3968\,\AA), \Mgb\ (5175\,\AA) and the
\caii\ triplet (8498, 8542, 8662\,\AA); see \eg\ \cite{Greene06}. A portion of  data are
shown in Figure \ref{fig_app_opt}, in which some of these absorption features can
clearly be seen. The continuum and the absorption features were fit using the
penalized PiXel Fitting software (pPXF; \citealt{Cappellari04}) after masking out all
of the prominent emission lines. The stellar features were fit using the VLT/Xshooter
stellar template library (\citealt{Chen14}), broadly following the procedure outlined in
\cite{Koss17}; these templates have been observed with a higher resolution ($R =
10,000$) than the observation of \eso, and were convolved to the appropriate
spectral resolution in pPXF. After accounting for the spectral resolution (which was
subtracted in quadrature), the data imply a velocity dispersion of $\sigma_{*} = 125
\pm 2$\,\kmps\ in the UVB arm and $\sigma_{*} = 125 \pm 3$\,\kmps\ in the VIS arm.
This implies a black hole mass of $\log[M_{\rm{BH}} / \msun] = 7.5 \pm 0.3$ from
the $M_{\rm{BH}} - \sigma_{*}$ relation, where the quoted uncertainty is dominated
by the known scatter of 0.3 dex around this trend.

\bibliographystyle{../../mnras}

\bibliography{../../references}

\begin{thebibliography}{155}
\expandafter\ifx\csname natexlab\endcsname\relax\def\natexlab#1{#1}\fi

\bibitem[{Alston} et~al.(2020){Alston}, {Fabian}, {Kara} et~al.]{Alston20iras}
{Alston} W.~N., {Fabian} A.~C., {Kara} E., et~al., 2020, Nature Astronomy, 4,
  597

\bibitem[{Arnaud}(1996)]{xspec}
{Arnaud} K.~A., 1996, in { Astronomical Data Analysis Software and Systems
  V\/}, edited by {G.~H.~Jacoby \& J.~Barnes}, vol. 101 of { Astron. Soc. Pac.
  Conference Series, Astron. Soc. Pac., San Francisco\/}, ~17

\bibitem[{Ballantyne} et~al.(2011){Ballantyne}, {McDuffie} \&
  {Rusin}]{Ballantyne11}
{Ballantyne} D.~R., {McDuffie} J.~R., {Rusin} J.~S., 2011, \apj, 734, 112

\bibitem[{Balokovi\'{c}}(2017)]{Balokovic17phd}
{Balokovi\'{c}} M., 2017, {Unveiling the Structure of Active Galactic Nuclei
  with Hard X-ray Spectroscopy}, Ph.D. thesis, California Institute of
  Technology

\bibitem[{Balokovi{\'c}} et~al.(2019){Balokovi{\'c}}, {Garc{\'\i}a} \&
  {Cabral}]{borus12}
{Balokovi{\'c}} M., {Garc{\'\i}a} J.~A., {Cabral} S.~E., 2019, Research Notes
  of the American Astronomical Society, 3, 11, 173

\bibitem[{Balokovi{\'c}} et~al.(2020){Balokovi{\'c}}, {Harrison}, {Madejski}
  et~al.]{Balokovic20}
{Balokovi{\'c}} M., {Harrison} F.~A., {Madejski} G., et~al., 2020, arXiv
  e-prints,  arXiv:2011.06583

\bibitem[{Balokovi{\'c}} et~al.(2015){Balokovi{\'c}}, {Matt}, {Harrison}
  et~al.]{Balokovic15}
{Balokovi{\'c}} M., {Matt} G., {Harrison} F.~A., et~al., 2015, \apj, 800, 1, 62

\bibitem[{Bardeen} et~al.(1972){Bardeen}, {Press} \& {Teukolsky}]{Bardeen72}
{Bardeen} J.~M., {Press} W.~H., {Teukolsky} S.~A., 1972, \apj, 178, 347

\bibitem[{Berti} \& {Volonteri}(2008)]{Berti08}
{Berti} E., {Volonteri} M., 2008, \apj, 684, 822

\bibitem[{Bianchi} et~al.(2005){Bianchi}, {Miniutti}, {Fabian} \&
  {Iwasawa}]{Bianchi05mrk3}
{Bianchi} S., {Miniutti} G., {Fabian} A.~C., {Iwasawa} K., 2005, \mnras, 360,
  1, 380

\bibitem[{Bird} et~al.(2007){Bird}, {Malizia}, {Bazzano} et~al.]{3ISGRI}
{Bird} A.~J., {Malizia} A., {Bazzano} A., et~al., 2007, \apjs, 170, 1, 175

\bibitem[{Blandford} \& {Znajek}(1977)]{BZ77}
{Blandford} R.~D., {Znajek} R.~L., 1977, \mnras, 179, 433

\bibitem[{Bonson} \& {Gallo}(2016)]{Bonson16}
{Bonson} K., {Gallo} L.~C., 2016, \mnras, 458, 1927

\bibitem[{Brenneman} et~al.(2014){Brenneman}, {Madejski}, {Fuerst}
  et~al.]{Brenneman14a}
{Brenneman} L.~W., {Madejski} G., {Fuerst} F., et~al., 2014, \apj, 781, 83

\bibitem[{Brenneman} et~al.(2011){Brenneman}, {Reynolds}, {Nowak}
  et~al.]{Brenneman11}
{Brenneman} L.~W., {Reynolds} C.~S., {Nowak} M.~A., et~al., 2011, \apj, 736,
  103

\bibitem[{Brightman} et~al.(2013){Brightman}, {Silverman}, {Mainieri}
  et~al.]{Brightman13}
{Brightman} M., {Silverman} J.~D., {Mainieri} V., et~al., 2013, \mnras, 433,
  2485

\bibitem[{Buisson} et~al.(2018){Buisson}, {Fabian} \&
  {Lohfink}]{Buisson18cutoff}
{Buisson} D.~J.~K., {Fabian} A.~C., {Lohfink} A.~M., 2018, \mnras, 481, 4, 4419

\bibitem[{Cappellari} \& {Emsellem}(2004)]{Cappellari04}
{Cappellari} M., {Emsellem} E., 2004, \pasp, 116, 816, 138

\bibitem[{Chamani} et~al.(2020){Chamani}, {Koljonen} \&
  {Savolainen}]{Chamani20}
{Chamani} W., {Koljonen} K., {Savolainen} T., 2020, \aap, 635, A172

\bibitem[{Chartas} et~al.(2009){Chartas}, {Kochanek}, {Dai}, {Poindexter} \&
  {Garmire}]{Chartas09}
{Chartas} G., {Kochanek} C.~S., {Dai} X., {Poindexter} S., {Garmire} G., 2009,
  \apj, 693, 174

\bibitem[{Chen} et~al.(2014){Chen}, {Trager}, {Peletier} et~al.]{Chen14}
{Chen} Y.-P., {Trager} S.~C., {Peletier} R.~F., et~al., 2014, \aap, 565, A117

\bibitem[{Choudhury} et~al.(2017){Choudhury}, {Garc{\'\i}a}, {Steiner} \&
  {Bambi}]{Choudhury17}
{Choudhury} K., {Garc{\'\i}a} J.~A., {Steiner} J.~F., {Bambi} C., 2017, \apj,
  851, 1, 57

\bibitem[{Dadina} \& {Cappi}(2004)]{Dadina04}
{Dadina} M., {Cappi} M., 2004, \aap, 413, 921

\bibitem[{Dai} et~al.(2010){Dai}, {Kochanek}, {Chartas} et~al.]{Dai10}
{Dai} X., {Kochanek} C.~S., {Chartas} G., et~al., 2010, \apj, 709, 278

\bibitem[{Dauser} et~al.(2014){Dauser}, {Garc{\'{\i}}a}, {Parker}, {Fabian} \&
  {Wilms}]{Dauser14}
{Dauser} T., {Garc{\'{\i}}a} J., {Parker} M.~L., {Fabian} A.~C., {Wilms} J.,
  2014, \mnras, 444, L100

\bibitem[{Dauser} et~al.(2016){Dauser}, {Garc{\'{\i}}a}, {Walton}
  et~al.]{relxill_norm}
{Dauser} T., {Garc{\'{\i}}a} J., {Walton} D.~J., et~al., 2016, \aap, 590, A76

\bibitem[{Dauser} et~al.(2013){Dauser}, {Garcia}, {Wilms} et~al.]{Dauser13}
{Dauser} T., {Garcia} J., {Wilms} J., et~al., 2013, \mnras, 430, 1694

\bibitem[{Dauser} et~al.(2010){Dauser}, {Wilms}, {Reynolds} \&
  {Brenneman}]{relconv}
{Dauser} T., {Wilms} J., {Reynolds} C.~S., {Brenneman} L.~W., 2010, \mnras,
  409, 1534

\bibitem[{de Grijp} et~al.(1992){de Grijp}, {Keel}, {Miley}, {Goudfrooij} \&
  {Lub}]{deGrijp92}
{de Grijp} M.~H.~K., {Keel} W.~C., {Miley} G.~K., {Goudfrooij} P., {Lub} J.,
  1992, \aaps, 96, 389

\bibitem[{Di Matteo} et~al.(2005){Di Matteo}, {Springel} \&
  {Hernquist}]{DiMatteo05nat}
{Di Matteo} T., {Springel} V., {Hernquist} L., 2005, \nat, 433, 604

\bibitem[{Dov{\v{c}}iak} \& {Done}(2016)]{Dovciak16}
{Dov{\v{c}}iak} M., {Done} C., 2016, Astronomische Nachrichten, 337, 4-5, 441

\bibitem[{Ebrero} et~al.(2016){Ebrero}, {Kriss}, {Kaastra} \& {Ely}]{Ebrero16}
{Ebrero} J., {Kriss} G.~A., {Kaastra} J.~S., {Ely} J.~C., 2016, \aap, 586, A72

\bibitem[{Elvis} et~al.(1994){Elvis}, {Wilkes}, {McDowell} et~al.]{Elvis94}
{Elvis} M., {Wilkes} B.~J., {McDowell} J.~C., et~al., 1994, \apjs, 95, 1

\bibitem[{Fabian}(2012)]{Fabian12rev}
{Fabian} A.~C., 2012, \araa, 50, 455

\bibitem[{Fabian} et~al.(2013){Fabian}, {Kara}, {Walton} et~al.]{Fabian13iras}
{Fabian} A.~C., {Kara} E., {Walton} D.~J., et~al., 2013, \mnras, 429, 2917

\bibitem[{Fabian} et~al.(2017){Fabian}, {Lohfink}, {Belmont}, {Malzac} \&
  {Coppi}]{Fabian17}
{Fabian} A.~C., {Lohfink} A., {Belmont} R., {Malzac} J., {Coppi} P., 2017,
  \mnras, 467, 3, 2566

\bibitem[{Fabian} et~al.(2015){Fabian}, {Lohfink}, {Kara}, {Parker},
  {Vasudevan} \& {Reynolds}]{Fabian15}
{Fabian} A.~C., {Lohfink} A., {Kara} E., {Parker} M.~L., {Vasudevan} R.,
  {Reynolds} C.~S., 2015, \mnras, 451, 4375

\bibitem[{Fabian} et~al.(1989){Fabian}, {Rees}, {Stella} \& {White}]{Fabian89}
{Fabian} A.~C., {Rees} M.~J., {Stella} L., {White} N.~E., 1989, \mnras, 238,
  729

\bibitem[{Fabian} et~al.(2009){Fabian}, {Zoghbi}, {Ross} et~al.]{FabZog09}
{Fabian} A.~C., {Zoghbi} A., {Ross} R.~R., et~al., 2009, \nat, 459, 540

\bibitem[{Fabian} et~al.(2012){Fabian}, {Zoghbi}, {Wilkins} et~al.]{Fabian12}
{Fabian} A.~C., {Zoghbi} A., {Wilkins} D., et~al., 2012, \mnras, 419, 116

\bibitem[{Ferrarese} et~al.(2006){Ferrarese}, {C{\^o}t{\'e}}, {Dalla Bont{\`a}}
  et~al.]{Ferrarese06}
{Ferrarese} L., {C{\^o}t{\'e}} P., {Dalla Bont{\`a}} E., et~al., 2006, \apjl,
  644, L21

\bibitem[{Fiacconi} et~al.(2018){Fiacconi}, {Sijacki} \& {Pringle}]{Fiacconi18}
{Fiacconi} D., {Sijacki} D., {Pringle} J.~E., 2018, \mnras

\bibitem[{Foreman-Mackey}(2016)]{corner}
{Foreman-Mackey} D., 2016, The Journal of Open Source Software, 1, 24

\bibitem[{Gallo} et~al.(2021){Gallo}, {Gonzalez} \& {Miller}]{Gallo21}
{Gallo} L.~C., {Gonzalez} A.~G., {Miller} J.~M., 2021, \apjl, 908, 2, L33

\bibitem[{Garc{\'{\i}}a} et~al.(2014){Garc{\'{\i}}a}, {Dauser}, {Lohfink}
  et~al.]{relxill}
{Garc{\'{\i}}a} J., {Dauser} T., {Lohfink} A., et~al., 2014, \apj, 782, 76

\bibitem[{Garc{\'{\i}}a} \& {Kallman}(2010)]{xillver}
{Garc{\'{\i}}a} J., {Kallman} T.~R., 2010, \apj, 718, 695

\bibitem[{Garc{\'{\i}}a} et~al.(2015){Garc{\'{\i}}a}, {Dauser}, {Steiner},
  {McClintock}, {Keck} \& {Wilms}]{Garcia15}
{Garc{\'{\i}}a} J.~A., {Dauser} T., {Steiner} J.~F., {McClintock} J.~E., {Keck}
  M.~L., {Wilms} J., 2015, \apjl, 808, L37

\bibitem[{Garc{\'\i}a} et~al.(2018){Garc{\'\i}a}, {Kallman}, {Bautista}
  et~al.]{Garcia18}
{Garc{\'\i}a} J.~A., {Kallman} T.~R., {Bautista} M., et~al., 2018, in {
  Workshop on Astrophysical Opacities\/}, vol. 515 of { Astronomical Society of
  the Pacific Conference Series\/},  282

\bibitem[{Gehrels} et~al.(2004){Gehrels}, {Chincarini}, {Giommi} et~al.]{SWIFT}
{Gehrels} N., {Chincarini} G., {Giommi} P., et~al., 2004, \apj, 611, 1005

\bibitem[{George} \& {Fabian}(1991)]{George91}
{George} I.~M., {Fabian} A.~C., 1991, \mnras, 249, 352

\bibitem[Geweke(1992)]{Geweke92}
Geweke J., 1992, in { Bayesian Statistics 4 (ed. J. M. Bernardo, J. O. Berger,
  A. P. Dawid and A. F. M. Smith)\/},  169--193, Clarendon Press, Oxford

\bibitem[{Gonzalez} et~al.(2017){Gonzalez}, {Wilkins} \& {Gallo}]{Gonzalez17}
{Gonzalez} A.~G., {Wilkins} D.~R., {Gallo} L.~C., 2017, \mnras, 472, 2, 1932

\bibitem[{Goodman} \& {Weare}(2010)]{MCMC_GW}
{Goodman} J., {Weare} J., 2010, Communications in Applied Mathematics and
  Computational Science, 5, 1, 65

\bibitem[{Greene} \& {Ho}(2006)]{Greene06}
{Greene} J.~E., {Ho} L.~C., 2006, \apj, 641, 1, 117

\bibitem[{Grevesse} \& {Sauval}(1998)]{Grevesse98}
{Grevesse} N., {Sauval} A.~J., 1998, \ssr, 85, 161

\bibitem[{Haardt} \& {Maraschi}(1991)]{Haardt91}
{Haardt} F., {Maraschi} L., 1991, \apjl, 380, L51

\bibitem[{Harrison} et~al.(2013){Harrison}, {Craig}, {Christensen}
  et~al.]{NUSTAR}
{Harrison} F.~A., {Craig} W.~W., {Christensen} F.~E., et~al., 2013, \apj, 770,
  103

\bibitem[{HI4PI Collaboration} et~al.(2016){HI4PI Collaboration}, {Ben Bekhti},
  {Fl{\"o}er} et~al.]{NH2016}
{HI4PI Collaboration}, {Ben Bekhti} N., {Fl{\"o}er} L., et~al., 2016, \aap,
  594, A116

\bibitem[{Hopkins} \& {Elvis}(2010)]{Hopkins10}
{Hopkins} P.~F., {Elvis} M., 2010, \mnras, 401, 7

\bibitem[{Ingram} et~al.(2019){Ingram}, {Mastroserio}, {Dauser}, {Hovenkamp},
  {van der Klis} \& {Garc{\'\i}a}]{Ingram19}
{Ingram} A., {Mastroserio} G., {Dauser} T., {Hovenkamp} P., {van der Klis} M.,
  {Garc{\'\i}a} J.~A., 2019, \mnras, 488, 1, 324

\bibitem[{Jansen} et~al.(2001){Jansen}, {Lumb}, {Altieri} et~al.]{XMM}
{Jansen} F., {Lumb} D., {Altieri} B., et~al., 2001, \aap, 365, L1

\bibitem[{Jiang} et~al.(2019){Jiang}, {Fabian}, {Dauser} et~al.]{Jiang19agn}
{Jiang} J., {Fabian} A.~C., {Dauser} T., et~al., 2019, \mnras, 489, 3, 3436

\bibitem[{Jiang} et~al.(2018){Jiang}, {Parker}, {Fabian} et~al.]{Jiang18iras}
{Jiang} J., {Parker} M.~L., {Fabian} A.~C., et~al., 2018, \mnras

\bibitem[{Kaastra} et~al.(2014){Kaastra}, {Kriss}, {Cappi} et~al.]{Kaastra14}
{Kaastra} J.~S., {Kriss} G.~A., {Cappi} M., et~al., 2014, Science, 345, 6192,
  64

\bibitem[{Kallman} \& {Bautista}(2001)]{xstar}
{Kallman} T., {Bautista} M., 2001, \apjs, 133, 221

\bibitem[{Kallman} et~al.(2021){Kallman}, {Bautista}, {Deprince}
  et~al.]{Kallman21}
{Kallman} T., {Bautista} M., {Deprince} J., et~al., 2021, \apj, 908, 1, 94

\bibitem[{Kammoun} et~al.(2019){Kammoun}, {Dom{\v{c}}ek}, {Svoboda},
  {Dov{\v{c}}iak} \& {Matt}]{Kammoun19}
{Kammoun} E.~S., {Dom{\v{c}}ek} V., {Svoboda} J., {Dov{\v{c}}iak} M., {Matt}
  G., 2019, \mnras, 485, 1, 239

\bibitem[{Kammoun} et~al.(2018){Kammoun}, {Nardini} \& {Risaliti}]{Kammoun18}
{Kammoun} E.~S., {Nardini} E., {Risaliti} G., 2018, \aap, 614, A44

\bibitem[{Kamraj} et~al.(2018){Kamraj}, {Harrison}, {Balokovi{\'c}}, {Lohfink}
  \& {Brightman}]{Kamraj18}
{Kamraj} N., {Harrison} F.~A., {Balokovi{\'c}} M., {Lohfink} A., {Brightman}
  M., 2018, \apj, 866, 2, 124

\bibitem[{Kara} et~al.(2016){Kara}, {Alston}, {Fabian} et~al.]{Kara16}
{Kara} E., {Alston} W.~N., {Fabian} A.~C., et~al., 2016, \mnras, 462, 1, 511

\bibitem[{King} \& {Pringle}(2006)]{King06}
{King} A.~R., {Pringle} J.~E., 2006, \mnras, 373, L90

\bibitem[{Kormendy} \& {Ho}(2013)]{Kormendy13rev}
{Kormendy} J., {Ho} L.~C., 2013, \araa, 51, 511

\bibitem[{Koss} et~al.(2017){Koss}, {Trakhtenbrot}, {Ricci} et~al.]{Koss17}
{Koss} M., {Trakhtenbrot} B., {Ricci} C., et~al., 2017, \apj, 850, 1, 74

\bibitem[{Laha} et~al.(2014){Laha}, {Guainazzi}, {Dewangan}, {Chakravorty} \&
  {Kembhavi}]{Laha14}
{Laha} S., {Guainazzi} M., {Dewangan} G.~C., {Chakravorty} S., {Kembhavi}
  A.~K., 2014, \mnras, 441, 3, 2613

\bibitem[{Lanz} et~al.(2019){Lanz}, {Hickox}, {Balokovi{\'c}} et~al.]{Lanz19}
{Lanz} L., {Hickox} R.~C., {Balokovi{\'c}} M., et~al., 2019, \apj, 870, 1, 26

\bibitem[{Laor}(1991)]{kdblur}
{Laor} A., 1991, \apj, 376, 90

\bibitem[{Liedahl} et~al.(1995){Liedahl}, {Osterheld} \& {Goldstein}]{mekal2}
{Liedahl} D.~A., {Osterheld} A.~L., {Goldstein} W.~H., 1995, \apjl, 438, L115

\bibitem[{Lusso} et~al.(2010){Lusso}, {Comastri}, {Vignali} et~al.]{Lusso10}
{Lusso} E., {Comastri} A., {Vignali} C., et~al., 2010, \aap, 512, A34

\bibitem[{MacLeod} et~al.(2015){MacLeod}, {Morgan}, {Mosquera}
  et~al.]{MacLeod15}
{MacLeod} C.~L., {Morgan} C.~W., {Mosquera} A., et~al., 2015, \apj, 806, 2, 258

\bibitem[{Madsen} et~al.(2020){Madsen}, {Grefenstette}, {Pike}
  et~al.]{NuSTARmli}
{Madsen} K.~K., {Grefenstette} B.~W., {Pike} S., et~al., 2020, arXiv e-prints,
  arXiv:2005.00569

\bibitem[{Madsen} et~al.(2015){Madsen}, {Harrison}, {Markwardt}
  et~al.]{NUSTARcal}
{Madsen} K.~K., {Harrison} F.~A., {Markwardt} C.~B., et~al., 2015, \apjs, 220,
  8

\bibitem[{Magdziarz} \& {Zdziarski}(1995)]{pexrav}
{Magdziarz} P., {Zdziarski} A.~A., 1995, \mnras, 273, 837

\bibitem[{Makino} \& {ASTRO-C Team}(1987)]{GINGA}
{Makino} F., {ASTRO-C Team}, 1987, \aplett, 25, 223

\bibitem[{Malizia} et~al.(2014){Malizia}, {Molina}, {Bassani}
  et~al.]{Malizia14}
{Malizia} A., {Molina} M., {Bassani} L., et~al., 2014, \apjl, 782, 2, L25

\bibitem[{Malkan} et~al.(1998){Malkan}, {Gorjian} \& {Tam}]{Malkan98}
{Malkan} M.~A., {Gorjian} V., {Tam} R., 1998, \apjs, 117, 1, 25

\bibitem[{Marchesi} et~al.(2017){Marchesi}, {Tremblay}, {Ajello}
  et~al.]{Marchesi17}
{Marchesi} S., {Tremblay} L., {Ajello} M., et~al., 2017, \apj, 848, 1, 53

\bibitem[{Marinucci} et~al.(2014){Marinucci}, {Matt}, {Kara}
  et~al.]{Marinucci14}
{Marinucci} A., {Matt} G., {Kara} E., et~al., 2014, \mnras, 440, 2347

\bibitem[{Martocchia} et~al.(2000){Martocchia}, {Karas} \&
  {Matt}]{Martocchia00}
{Martocchia} A., {Karas} V., {Matt} G., 2000, \mnras, 312, 4, 817

\bibitem[{Matt} et~al.(2015){Matt}, {Balokovi{\'c}}, {Marinucci}
  et~al.]{Matt15}
{Matt} G., {Balokovi{\'c}} M., {Marinucci} A., et~al., 2015, \mnras, 447, 4,
  3029

\bibitem[{McHardy} et~al.(2006){McHardy}, {Koerding}, {Knigge}, {Uttley} \&
  {Fender}]{McHardy06}
{McHardy} I.~M., {Koerding} E., {Knigge} C., {Uttley} P., {Fender} R.~P., 2006,
  \nat, 444, 730

\bibitem[{Mehdipour} et~al.(2017){Mehdipour}, {Kaastra}, {Kriss}
  et~al.]{Mehdipour17}
{Mehdipour} M., {Kaastra} J.~S., {Kriss} G.~A., et~al., 2017, \aap, 607, A28

\bibitem[{Mewe} et~al.(1986){Mewe}, {Lemen} \& {van den Oord}]{mekal1}
{Mewe} R., {Lemen} J.~R., {van den Oord} G.~H.~J., 1986, \aaps, 65, 511

\bibitem[{Miniutti} \& {Fabian}(2004)]{lightbending}
{Miniutti} G., {Fabian} A.~C., 2004, \mnras, 349, 1435

\bibitem[{Miniutti} et~al.(2007){Miniutti}, {Fabian}, {Anabuki}
  et~al.]{Miniutti07}
{Miniutti} G., {Fabian} A.~C., {Anabuki} N., et~al., 2007, \pasj, 59, 315

\bibitem[{Moderski} et~al.(1998){Moderski}, {Sikora} \& {Lasota}]{Moderski98}
{Moderski} R., {Sikora} M., {Lasota} J.-P., 1998, \mnras, 301, 142

\bibitem[{Nandra} et~al.(2007){Nandra}, {O'Neill}, {George} \&
  {Reeves}]{Nandra07}
{Nandra} K., {O'Neill} P.~M., {George} I.~M., {Reeves} J.~N., 2007, \mnras,
  382, 194

\bibitem[{Nardini} et~al.(2015){Nardini}, {Reeves}, {Gofford}
  et~al.]{Nardini15}
{Nardini} E., {Reeves} J.~N., {Gofford} J., et~al., 2015, Science, 347, 860

\bibitem[{Oh} et~al.(2018){Oh}, {Koss}, {Markwardt} et~al.]{BAT105m}
{Oh} K., {Koss} M., {Markwardt} C.~B., et~al., 2018, \apjs, 235, 1, 4

\bibitem[{Panagiotou} \& {Walter}(2019)]{Panagiotou19}
{Panagiotou} C., {Walter} R., 2019, \aap, 626, A40

\bibitem[{Parker} et~al.(2014){Parker}, {Wilkins}, {Fabian}
  et~al.]{Parker14mrk}
{Parker} M.~L., {Wilkins} D.~R., {Fabian} A.~C., et~al., 2014, \mnras, 443,
  1723

\bibitem[{Peterson}(2014)]{Peterson14rev}
{Peterson} B.~M., 2014, \ssr, 183, 1-4, 253

\bibitem[{Petrucci} et~al.(2001){Petrucci}, {Merloni}, {Fabian}, {Haardt} \&
  {Gallo}]{Petrucci01}
{Petrucci} P.~O., {Merloni} A., {Fabian} A., {Haardt} F., {Gallo} E., 2001,
  \mnras, 328, 501

\bibitem[{Ponti} et~al.(2012){Ponti}, {Papadakis}, {Bianchi}
  et~al.]{Ponti12var}
{Ponti} G., {Papadakis} I., {Bianchi} S., et~al., 2012, \aap, 542, A83

\bibitem[{Reeves} et~al.(2018){Reeves}, {Braito}, {Nardini}, {Lobban}, {Matzeu}
  \& {Costa}]{Reeves18}
{Reeves} J.~N., {Braito} V., {Nardini} E., {Lobban} A.~P., {Matzeu} G.~A.,
  {Costa} M.~T., 2018, \apjl, 854, L8

\bibitem[{Reis} et~al.(2014){Reis}, {Reynolds}, {Miller} \&
  {Walton}]{Reis14nat}
{Reis} R.~C., {Reynolds} M.~T., {Miller} J.~M., {Walton} D.~J., 2014, \nat,
  507, 207

\bibitem[{Reynolds}(2020)]{Reynolds20rev}
{Reynolds} C.~S., 2020, arXiv e-prints,  arXiv:2011.08948

\bibitem[{Reynolds} \& {Begelman}(1997)]{Reynolds97feK}
{Reynolds} C.~S., {Begelman} M.~C., 1997, \apj, 488, 109

\bibitem[{Reynolds} et~al.(2012){Reynolds}, {Brenneman}, {Lohfink}
  et~al.]{Reynolds12}
{Reynolds} C.~S., {Brenneman} L.~W., {Lohfink} A.~M., et~al., 2012, \apj, 755,
  88

\bibitem[{Reynolds} et~al.(2014){Reynolds}, {Walton}, {Miller} \&
  {Reis}]{Reynolds14}
{Reynolds} M.~T., {Walton} D.~J., {Miller} J.~M., {Reis} R.~C., 2014, \apjl,
  792, L19

\bibitem[{Ricci} et~al.(2018){Ricci}, {Ho}, {Fabian} et~al.]{Ricci18}
{Ricci} C., {Ho} L.~C., {Fabian} A.~C., et~al., 2018, \mnras, 480, 2, 1819

\bibitem[{Ricci} et~al.(2017){Ricci}, {Trakhtenbrot}, {Koss} et~al.]{Ricci17}
{Ricci} C., {Trakhtenbrot} B., {Koss} M.~J., et~al., 2017, \apjs, 233, 2, 17

\bibitem[{Risaliti} et~al.(2005){Risaliti}, {Bianchi}, {Matt}
  et~al.]{Risaliti05b}
{Risaliti} G., {Bianchi} S., {Matt} G., et~al., 2005, \apjl, 630, L129

\bibitem[{Risaliti} et~al.(2013){Risaliti}, {Harrison}, {Madsen}
  et~al.]{Risaliti13nat}
{Risaliti} G., {Harrison} F.~A., {Madsen} K.~K., et~al., 2013, \nat, 494, 449

\bibitem[{Risaliti} et~al.(2009{\natexlab{a}}){Risaliti}, {Miniutti}, {Elvis}
  et~al.]{Risaliti09a}
{Risaliti} G., {Miniutti} G., {Elvis} M., et~al., 2009{\natexlab{a}}, \apj,
  696, 160

\bibitem[{Risaliti} et~al.(2009{\natexlab{b}}){Risaliti}, {Young} \&
  {Elvis}]{Risaliti09LxGam}
{Risaliti} G., {Young} M., {Elvis} M., 2009{\natexlab{b}}, \apjl, 700, L6

\bibitem[{Rivers} et~al.(2015){Rivers}, {Risaliti}, {Walton} et~al.]{Rivers15}
{Rivers} E., {Risaliti} G., {Walton} D.~J., et~al., 2015, \apj, 804, 107

\bibitem[{Sesana} et~al.(2014){Sesana}, {Barausse}, {Dotti} \&
  {Rossi}]{Sesana14}
{Sesana} A., {Barausse} E., {Dotti} M., {Rossi} E.~M., 2014, \apj, 794, 104

\bibitem[{Shakura} \& {Sunyaev}(1973)]{Shakura73}
{Shakura} N.~I., {Sunyaev} R.~A., 1973, \aap, 24, 337

\bibitem[{Shemmer} et~al.(2008){Shemmer}, {Brandt}, {Netzer}, {Maiolino} \&
  {Kaspi}]{Shemmer08}
{Shemmer} O., {Brandt} W.~N., {Netzer} H., {Maiolino} R., {Kaspi} S., 2008,
  \apj, 682, 1, 81

\bibitem[{Sikora} et~al.(2007){Sikora}, {Stawarz} \& {Lasota}]{Sikora07}
{Sikora} M., {Stawarz} {\L}., {Lasota} J.-P., 2007, \apj, 658, 2, 815

\bibitem[{Smith} \& {Done}(1996)]{Smith96}
{Smith} D.~A., {Done} C., 1996, \mnras, 280, 2, 355

\bibitem[{Str{\"u}der} et~al.(2001){Str{\"u}der}, {Briel}, {Dennerl}
  et~al.]{XMM_PN}
{Str{\"u}der} L., {Briel} U., {Dennerl} K., et~al., 2001, \aap, 365, L18

\bibitem[{Svoboda} et~al.(2012){Svoboda}, {Dov{\v{c}}iak}, {Goosmann}
  et~al.]{Svoboda12}
{Svoboda} J., {Dov{\v{c}}iak} M., {Goosmann} R.~W., et~al., 2012, \aap, 545,
  A106

\bibitem[{Tanaka} et~al.(1994){Tanaka}, {Inoue} \& {Holt}]{ASCA}
{Tanaka} Y., {Inoue} H., {Holt} S.~S., 1994, \pasj, 46, L37

\bibitem[{Taylor} \& {Reynolds}(2018)]{fenrir}
{Taylor} C., {Reynolds} C.~S., 2018, \apj, 855, 120

\bibitem[{Tombesi} et~al.(2012){Tombesi}, {Cappi}, {Reeves} \&
  {Braito}]{Tombesi12}
{Tombesi} F., {Cappi} M., {Reeves} J.~N., {Braito} V., 2012, \mnras, 422, 1, L1

\bibitem[{Tomsick} et~al.(2018){Tomsick}, {Parker}, {Garc{\'{\i}}a}
  et~al.]{Tomsick18cyg}
{Tomsick} J.~A., {Parker} M.~L., {Garc{\'{\i}}a} J.~A., et~al., 2018, \apj,
  855, 3

\bibitem[{Tortosa} et~al.(2018){Tortosa}, {Bianchi}, {Marinucci}, {Matt} \&
  {Petrucci}]{Tortosa18}
{Tortosa} A., {Bianchi} S., {Marinucci} A., {Matt} G., {Petrucci} P.~O., 2018,
  \aap, 614, A37

\bibitem[{Trakhtenbrot} et~al.(2017){Trakhtenbrot}, {Ricci}, {Koss}
  et~al.]{Trakhtenbrot17}
{Trakhtenbrot} B., {Ricci} C., {Koss} M.~J., et~al., 2017, \mnras, 470, 1, 800

\bibitem[{Tueller} et~al.(2008){Tueller}, {Mushotzky}, {Barthelmy}
  et~al.]{BAT9m}
{Tueller} J., {Mushotzky} R.~F., {Barthelmy} S., et~al., 2008, \apj, 681, 1,
  113

\bibitem[{Turner} et~al.(2001){Turner}, {Abbey}, {Arnaud} et~al.]{XMM_MOS}
{Turner} M.~J.~L., {Abbey} A., {Arnaud} M., et~al., 2001, \aap, 365, L27

\bibitem[{Ursini} et~al.(2020){Ursini}, {Dov{\v{c}}iak}, {Zhang}, {Matt},
  {Petrucci} \& {Done}]{Ursini20}
{Ursini} F., {Dov{\v{c}}iak} M., {Zhang} W., {Matt} G., {Petrucci} P.~O.,
  {Done} C., 2020, \aap, 644, A132

\bibitem[{Vasudevan} \& {Fabian}(2009)]{Vasudevan09}
{Vasudevan} R.~V., {Fabian} A.~C., 2009, \mnras, 392, 1124

\bibitem[{Vasudevan} et~al.(2016){Vasudevan}, {Fabian}, {Reynolds}, {Aird},
  {Dauser} \& {Gallo}]{Vasudevan16}
{Vasudevan} R.~V., {Fabian} A.~C., {Reynolds} C.~S., {Aird} J., {Dauser} T.,
  {Gallo} L.~C., 2016, \mnras, 458, 2012

\bibitem[{Vaughan} et~al.(2003){Vaughan}, {Fabian} \& {Nandra}]{Vaughan03}
{Vaughan} S., {Fabian} A.~C., {Nandra} K., 2003, \mnras, 339, 1237

\bibitem[{Verner} et~al.(1996){Verner}, {Ferland}, {Korista} \&
  {Yakovlev}]{Verner96}
{Verner} D.~A., {Ferland} G.~J., {Korista} K.~T., {Yakovlev} D.~G., 1996, \apj,
  465, 487

\bibitem[{Vernet} et~al.(2011){Vernet}, {Dekker}, {D'Odorico} et~al.]{XSHOOTER}
{Vernet} J., {Dekker} H., {D'Odorico} S., et~al., 2011, \aap, 536, A105

\bibitem[{Vignali} et~al.(1998){Vignali}, {Comastri}, {Stirpe}
  et~al.]{Vignali98}
{Vignali} C., {Comastri} A., {Stirpe} G.~M., et~al., 1998, \aap, 333, 411

\bibitem[{Walton} et~al.(2020){Walton}, {Alston}, {Kosec} et~al.]{Walton20iras}
{Walton} D.~J., {Alston} W.~N., {Kosec} P., et~al., 2020, \mnras, 499, 1, 1480

\bibitem[{Walton} et~al.(2018){Walton}, {Brightman}, {Risaliti}
  et~al.]{Walton18}
{Walton} D.~J., {Brightman} M., {Risaliti} G., et~al., 2018, \mnras, 473, 4377

\bibitem[{Walton} et~al.(2013){Walton}, {Nardini}, {Fabian}, {Gallo} \&
  {Reis}]{Walton13spin}
{Walton} D.~J., {Nardini} E., {Fabian} A.~C., {Gallo} L.~C., {Reis} R.~C.,
  2013, \mnras, 428, 2901

\bibitem[{Walton} et~al.(2019){Walton}, {Nardini}, {Gallo} et~al.]{Walton19ufo}
{Walton} D.~J., {Nardini} E., {Gallo} L.~C., et~al., 2019, \mnras, 484, 2544

\bibitem[{Walton} et~al.(2014){Walton}, {Risaliti}, {Harrison}
  et~al.]{Walton14}
{Walton} D.~J., {Risaliti} G., {Harrison} F.~A., et~al., 2014, \apj, 788, 76

\bibitem[{Walton} et~al.(2016){Walton}, {Tomsick}, {Madsen}
  et~al.]{Walton16cyg}
{Walton} D.~J., {Tomsick} J.~A., {Madsen} K.~K., et~al., 2016, \apj, 826, 87

\bibitem[{Weisskopf} et~al.(2002){Weisskopf}, {Brinkman}, {Canizares},
  {Garmire}, {Murray} \& {Van Speybroeck}]{CHANDRA}
{Weisskopf} M.~C., {Brinkman} B., {Canizares} C., {Garmire} G., {Murray} S.,
  {Van Speybroeck} L.~P., 2002, \pasp, 114, 1

\bibitem[{Wilkins} \& {Fabian}(2011)]{Wilkins11}
{Wilkins} D.~R., {Fabian} A.~C., 2011, \mnras, 414, 1269

\bibitem[{Wilkins} \& {Fabian}(2012)]{Wilkins12}
{Wilkins} D.~R., {Fabian} A.~C., 2012, \mnras, 424, 1284

\bibitem[{Wilms} et~al.(2000){Wilms}, {Allen} \& {McCray}]{tbabs}
{Wilms} J., {Allen} A., {McCray} R., 2000, \apj, 542, 914

\bibitem[{Wilson} \& {Colbert}(1995)]{Wilson95}
{Wilson} A.~S., {Colbert} E.~J.~M., 1995, \apj, 438, 62

\bibitem[{Winkler} et~al.(2003){Winkler}, {Courvoisier}, {Di Cocco}
  et~al.]{INTEGRAL}
{Winkler} C., {Courvoisier} T.~J.~L., {Di Cocco} G., et~al., 2003, \aap, 411,
  L1

\bibitem[{Winter} et~al.(2008){Winter}, {Mushotzky}, {Tueller} \&
  {Markwardt}]{Winter08}
{Winter} L.~M., {Mushotzky} R.~F., {Tueller} J., {Markwardt} C., 2008, \apj,
  674, 2, 686

\bibitem[{Zdziarski} et~al.(1996){Zdziarski}, {Johnson} \&
  {Magdziarz}]{nthcomp1}
{Zdziarski} A.~A., {Johnson} W.~N., {Magdziarz} P., 1996, \mnras, 283, 193

\bibitem[{Zhang} et~al.(2019){Zhang}, {Dov{\v{c}}iak} \& {Bursa}]{Zhang19}
{Zhang} W., {Dov{\v{c}}iak} M., {Bursa} M., 2019, \apj, 875, 2, 148

\bibitem[{Zoghbi} et~al.(2012){Zoghbi}, {Fabian}, {Reynolds} \&
  {Cackett}]{Zoghbi12}
{Zoghbi} A., {Fabian} A.~C., {Reynolds} C.~S., {Cackett} E.~M., 2012, \mnras,
  422, 129

\bibitem[{Zycki} et~al.(1999){Zycki}, {Done} \& {Smith}]{nthcomp2}
{Zycki} P.~T., {Done} C., {Smith} D.~A., 1999, \mnras, 309, 561

\end{thebibliography}

\label{lastpage}

\end{document}